\documentclass[conference]{IEEEtran}
\IEEEoverridecommandlockouts
\usepackage{cite}
\usepackage{amsmath,amssymb,amsfonts}
\usepackage{algorithmic}
\usepackage{graphicx}
\usepackage{textcomp}
\usepackage{xcolor}
\usepackage{verbatim}
\usepackage{hyperref}
\usepackage{threeparttable}
\usepackage{array}
\usepackage{booktabs}
\usepackage{multirow}
\usepackage{xspace}
\usepackage{ulem}
\usepackage{epigraph}
\def\BibTeX{{\rm B\kern-.05em{\sc i\kern-.025em b}\kern-.08em
    T\kern-.1667em\lower.7ex\hbox{E}\kern-.125emX}}

\newcommand{\tool}{{\texttt{GUIGAN}}\xspace}
\newcommand{\revised}[1]{{#1}}

\begin{document}

\title{\tool: Learning to Generate GUI Designs Using Generative Adversarial Networks}

\author{\IEEEauthorblockN{Tianming Zhao}
\IEEEauthorblockA{\textit{Jilin University} \\
Changchun, China \\
zhaotm16@mails.jlu.edu.cn}
\and
\IEEEauthorblockN{Chunyang Chen\IEEEauthorrefmark{1}}
\IEEEauthorblockA{\textit{Monash University} \\
Melbourne, Australia \\
Chunyang.Chen@monash.edu}
\and 
\IEEEauthorblockN{Yuanning Liu}
\IEEEauthorblockA{\textit{Jilin University} \\
Changchun, China \\
lyn@jlu.edu.cn}
\and
\IEEEauthorblockN{Xiaodong Zhu\IEEEauthorrefmark{1}}
\IEEEauthorblockA{\textit{Jilin University} \\
Changchun, China \\
zhuxd@jlu.edu.cn}
\thanks{* Corresponding author.}
}

\maketitle

\begin{abstract}
Graphical User Interface (GUI) is ubiquitous in almost all modern desktop software, mobile applications, and online websites.
A good GUI design is crucial to the success of the software in the market, but designing a good GUI which requires much innovation and creativity is difficult even to well-trained designers.
Besides, the requirement of the rapid development of GUI design also aggravates designers' working load.
So, the availability of various automated generated GUIs can help enhance the design personalization and specialization as they can cater to the taste of different designers.
To assist designers, we develop a model \tool to automatically generate GUI designs.
Different from conventional image generation models based on image pixels, our \tool is to reuse GUI components collected from existing mobile app GUIs for composing a new design that is similar to natural-language generation.
Our \tool is based on SeqGAN by modeling the GUI component style compatibility and GUI structure.
The evaluation demonstrates that our model significantly outperforms the best of the baseline methods by 30.77\% in Fr\'echet Inception distance (FID) and 12.35\% in 1-Nearest Neighbor Accuracy (1-NNA).
Through a pilot user study, we provide initial evidence of the usefulness of our approach for generating acceptable brand new GUI designs.
\end{abstract}


\begin{figure*}
  \includegraphics[width=\textwidth]{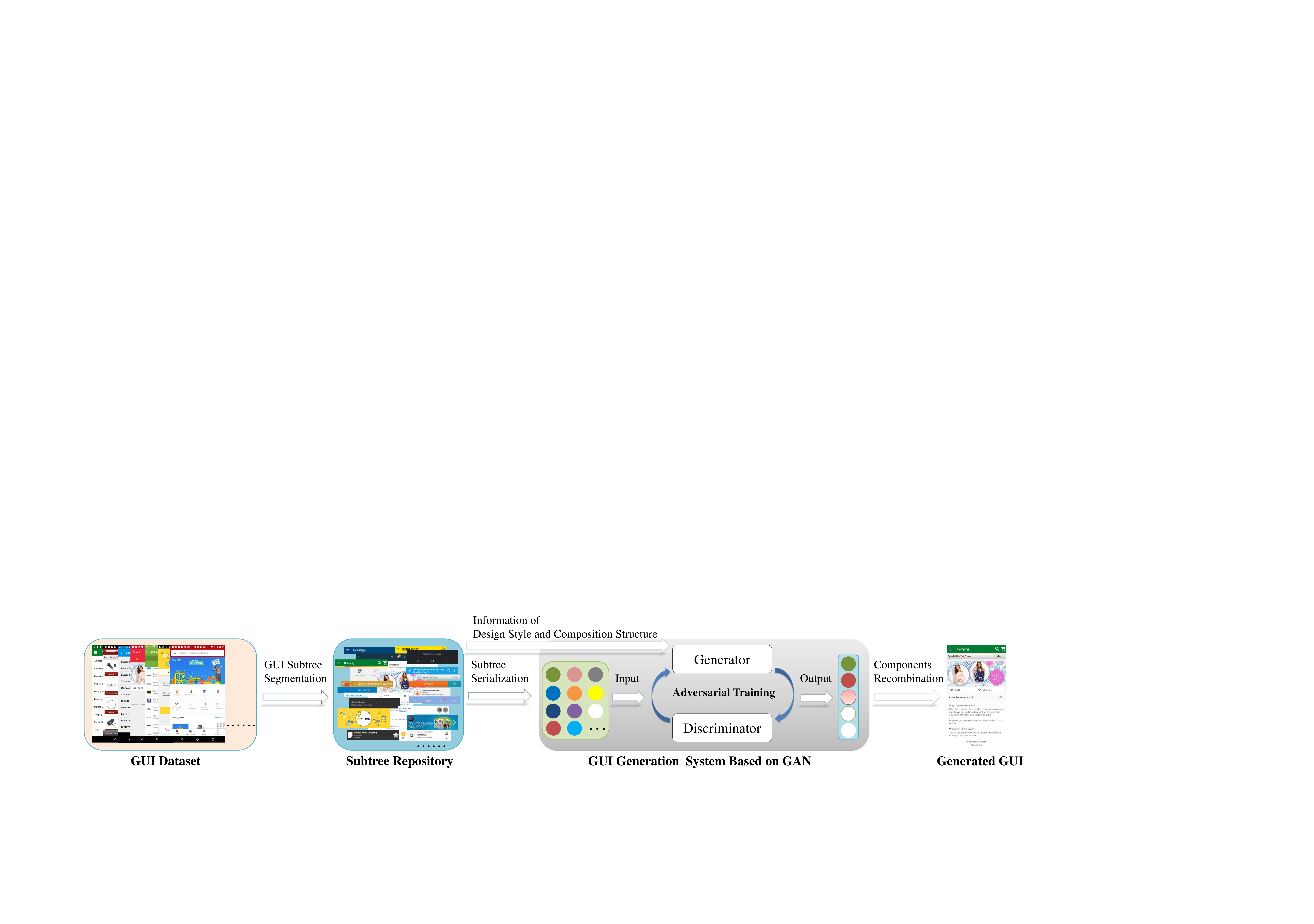}
  \caption{Overview of the proposed method.}
  %\Description{Overview of the proposed method.}
  \label{fig:overview}
\end{figure*}

\section{Introduction}
%BACKGROUND
Graphic User Interface (GUI) is ubiquitous in almost all modern desktop software, mobile applications, and online websites.
It provides a visual bridge between a software application and end-users through which they can interact with each other.
A good GUI design makes an application easy, practical, and efficient to use, which significantly affects the success of the application and the loyalty of its users~\cite{jansen1998graphical}.
For example, computer users view Apple's Macintosh system as having a better GUI than the Windows system, therefore their positive views almost double that of Windows users, leading to 20\% more brand loyalty~\cite{winograd1995programming}.

%FACT that developers also need to carry out the UI design
Good GUI design is difficult and time-consuming, even for professional GUI designers, as the designing process must follow many design rules and principles~\cite{web:appleDesign}~\cite{web:googlematerial}, such as fluent interactivity, universal usability, clear readability, aesthetic appearance, and consistent styles~\cite{galitz2007essential}~\cite{clifton2015android}~\cite{web:appleDesign}.
To follow the fashion trend, GUI designers have to keep reviewing the latest/hottest mobile apps/software or getting inspiration from design sharing sites (e.g., Dribbble\footnote{\url{https://dribbble.com/}}). %sites\footnote{\url{https://github.com/GUIDesignResearch/GUIGAN}}.
Considering that each mobile app/website/software contains tens of different screens and their GUIs need to be updated iteratively due to the market pressure, designers have to take much innovation-extensive working load.

Unfortunately, this design work often awaits just very few designers in a company~\cite{hong2011matters}, and software developers have to fill in the gap.
In a survey of more than 5,700 developers~\cite{web:developerDesign}, 51\% developers reported working on app GUI design tasks, more so than other development tasks, which they had to perform every few days.
However, software developers often do not have sufficient professional UI/UX design training with art sense.
That is why it is challenging for developers to design the GUI only from scratch.
Instead, when designing the GUI for websites or mobile apps, developers are very likely to search existing GUI designs on the internet as the reference, and further implement and customize the GUI design for their own purposes~\cite{web:wsdesign, web:startbootstrap}. 
This process usually happens at GUI development in small start-ups or small-scale open-source software, as there are not any professional UI/UX designers.

Although some studies help with the GUI search by attribute filtering~\cite{bernal2019guigle} or parsing code structure of UI~\cite{behrang2018guifetch}, there are three problems with the GUI search.
First, there is a gap between the developers' intention in mind and the output textual query, and another gap between the textual query and visual GUI design.
Due to the gap between the textual and visual information, the retrieved GUI may not satisfy developers' requirements.
Second, the retrieved GUI design may be adopted by other developers, resulting in the high similarity to other apps, negatively influencing the uniqueness of the app.
Directly using others' GUI may also involve potential intellectual property issues.
Third, the design style of some retrieved GUIs may be out of date and it's hard for developers to keep track of the latest trend of the GUI design.

\epigraph{``\textit{A lot of times, people don't know what they want until you show it to them.}''}{--- Steve Jobs}

%``\textit{A lot of times, people don't know what they want until you show it to them.}''--- Steve Jobs

Consequently, an automated method for creative GUI design generation is terribly needed to alleviate the burden of both novice designers and developers. 
With the generated GUI design, developers can further adopt the automated GUI code generation ~\cite{chen2018ui, beltramelli2018pix2code, moran2018machine} for the automated.
In that way, the overall GUI development process will be significantly simplified.

In this work, we develop a deep learning model \tool to automatically generate GUI designs based on the existing GUI screenshots collected from thousands of mobile apps.
It can provide designers and developers brand new GUI designs, and they can further customize the generated GUI for their own purpose, rather than starting from scratch.
Although there are plenty of image generation models like DCGAN, VAE-GAN, CycleGan, and WGAN~\cite{radford2015unsupervised, larsen2016autoencoding, zhu2017unpaired, arjovsky2017wasserstein}, they are all based on plain pixels.
In contrast, GUI is composed of a set of detailed components (e.g., button, text, images), and a good GUI design is concerned more with the composition of these components, rather than fine-grained component pixels.
Due to the characteristic of GUI and inspired by the natural-language generation (i.e., selecting a list of words for composing one sentence), we formulate our task as selecting a list of existing GUI component subtree to compose new GUI designs.

An overview of our approach is shown in Fig~\ref{fig:overview}.
First, We collect 12,230 GUI screenshots and their corresponding meta-information from 1,609 Android apps in 27 categories from Google Play and decompose them into 41,813 component subtrees for re-using.
Second, we develop a SeqGAN~\cite{yu2017seqgan} based model.
Apart from the default generation and discrimination loss, we model the GUI component style compatibility and GUI layout structure for guiding the training.
Therefore, our \tool can generate brand new GUI designs for designers' inspiration. 
The evaluation demonstrates that our model significantly outperforms the best of the baseline methods by 30.77\% in Fr\'echet Inception distance (FID) and 12.35\% in 1-Nearest Neighbor Accuracy (1-NNA).
Through a pilot user study, we provide the initial evidence of the usefulness of our approach for generating acceptable brand new GUIs. 

Our contributions in this work can be summarized as follow:
\begin{itemize}
	\item To the best of our knowledge, this is the first study to automatically generate the mobile app GUI design which requires much creativity and visual understanding. 
	\item We propose a novel deep learning-based method to generate brand new GUI designs composed of subtree sequences from the existing GUI designs without additional manual presets. 
	\item The experimental results based on two specific development conditions show that our method can successfully capture GUI design styles and structural features, and automatically generate a new composite GUI that conforms to the aesthetic of the consumers and standard GUI structure.
\end{itemize}

\begin{figure}
	\centerline{\includegraphics[width=3.2in]{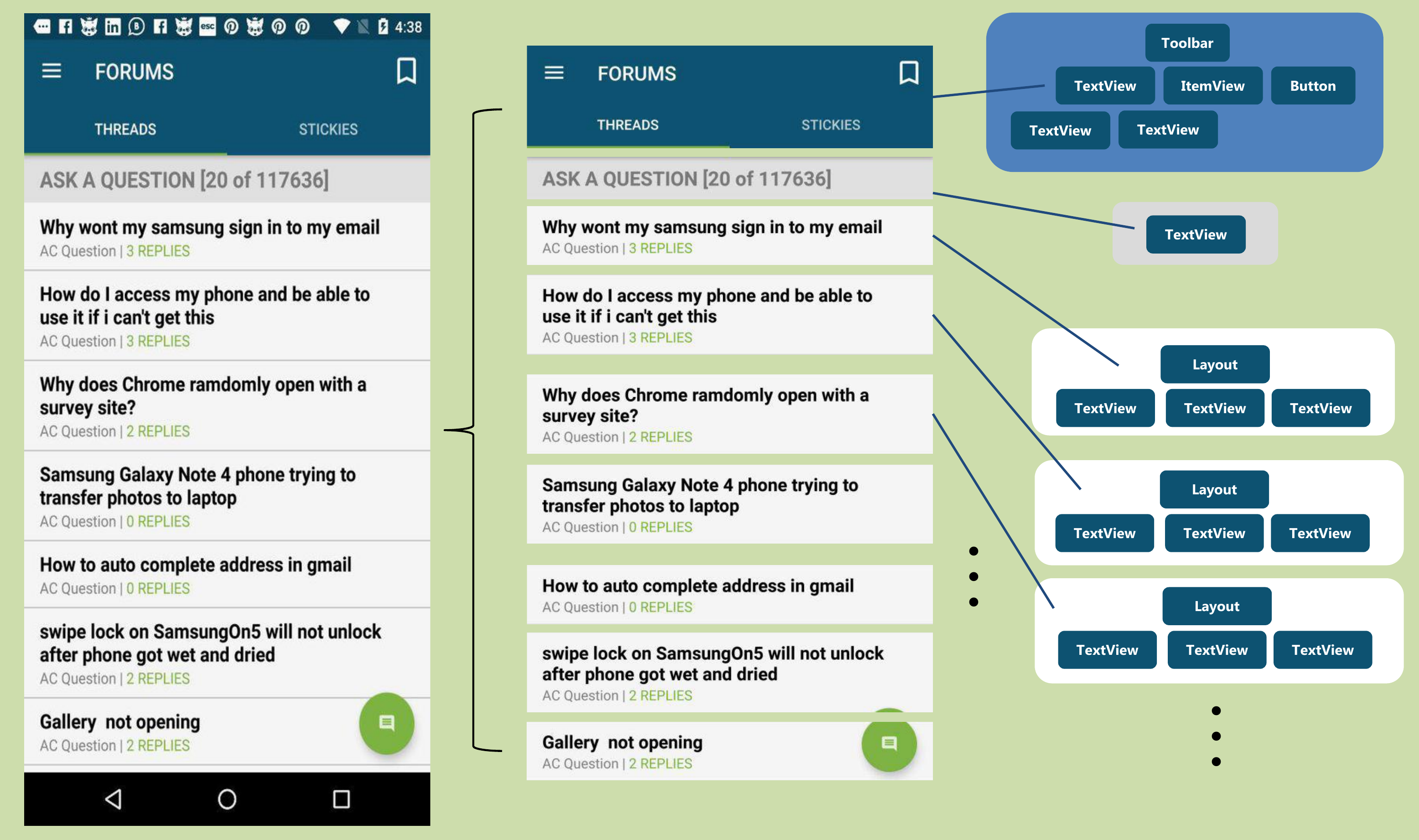}}
	\caption{Real-world data collection of GUI subtrees.}
	\label{fig:subtree}
\end{figure}

\begin{figure*} [h]
	\centerline{\includegraphics[width=7.2in]{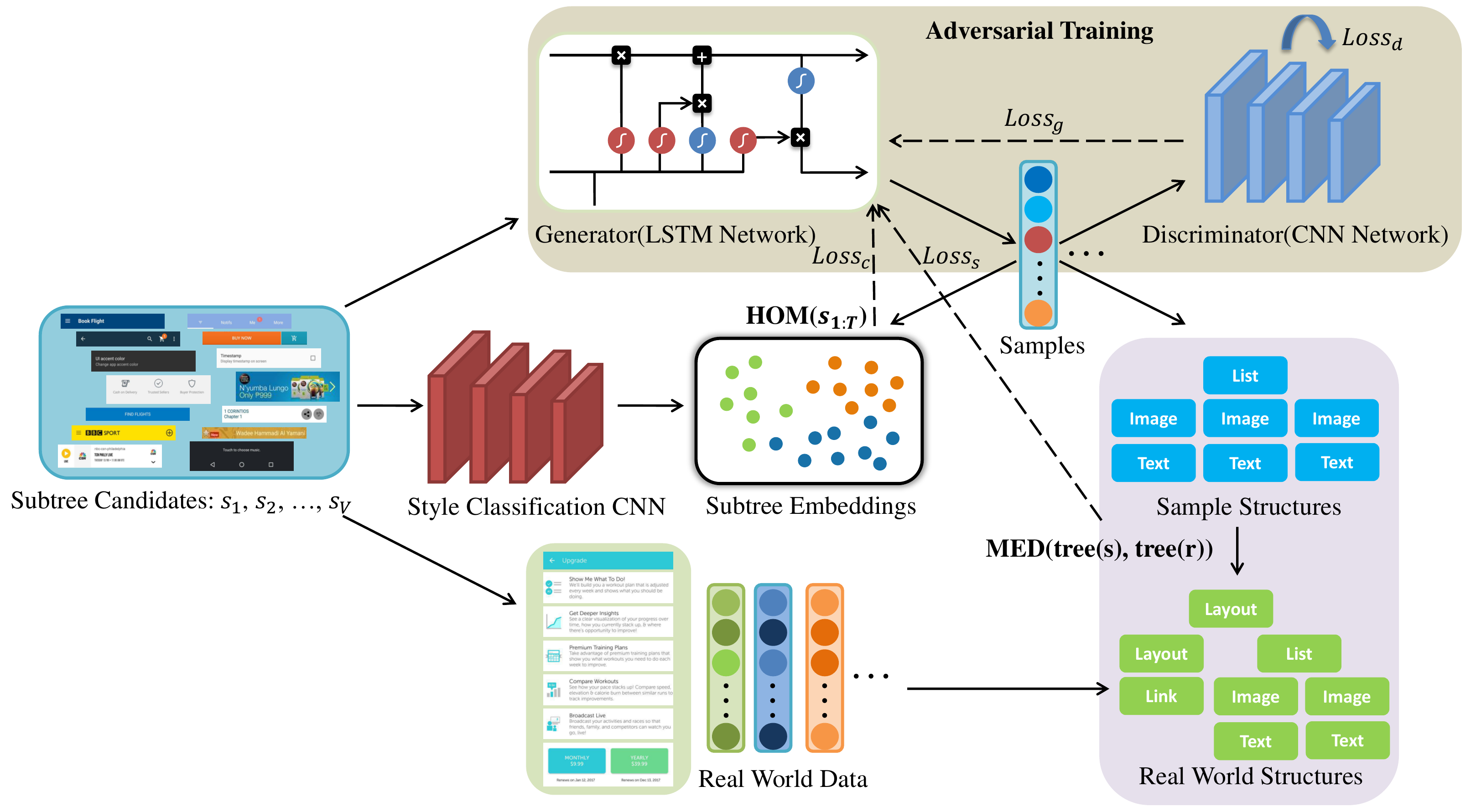}}
	\caption{The workflow of \tool.}
	\label{fig:approach}
\end{figure*}

\section{Preliminary}
In this section, we clear our goal and establish the corresponding task, and then introduce a deep learning method that our work is based on.

\subsection{Task Establishment}
\label{sec:task}
Different from the plain image which is made up of pixels, one GUI design image consists of two types of components i.e., widgets (e.g., button, image, text) and spatial layouts (e.g., linear layout, relative layout).
The widgets (leaf nodes) are organized by the layout (intermedia nodes) as the structural tree for one GUI design as seen in Fig~\ref{fig:subtree}. 
As most GUI designers may re-use some of their previous design components in their new design~\cite{zheng2019faceoff}, we take the subtree of existing GUIs as the basic unit for composing a new GUI design rather than plain pixels.
Therefore, we formulate our task as producing a sequence $S_{1:T} = (s_1,...,s_t,...s_T), s_t \in \textbf{S}$ of GUI component subtrees, where $\textbf{S}$ is the subtree repository. It can also be described as generating a new GUI by selecting a list of compatible GUI subtrees.

To obtain these candidate subtrees from screenshots of the GUIs, we cut them from the original screenshot according to certain rules.
Given one GUI design with detailed component information, we cut out all the first-level subtrees from the original DOM tree as seen in Fig~\ref{fig:subtree}. 
If the width of a subtree exceeds 90\% of the GUI width, we continue to cut it to the next level, otherwise, stop splitting and this subtree is used as the smallest granularity unit. The procedure will be iterated until the segmentation stops. Finally, \revised{all the smallest subtrees are given a unique number identification}. We remove the subtrees with duplicate bounds in one GUI and keep only one in the process. \revised{Based on the collection and observation of the data from our pilot study, we remove the subtrees with partial overlap and preserve those with the aspect ratio between 0.25 to 50, which has a specific structure and can be clipped from the original GUI screenshot.}

\subsection{Base Model}
Our work is mainly based on the generative adversarial networks (GAN)~\cite{goodfellow2014generative}, which consists of a generative network as the generator and a discriminative network as the discriminator respectively. The generator learns the features from the real data and generates new samples to fool the discriminator. The discriminator tries to distinguish the true sample from the fake one. These two networks are trained in an adversarial mode until the discriminator cannot distinguish the samples generated by the generator. 

The \tool is proposed based on SeqGAN~\cite{yu2017seqgan}, which is a variant of GAN. SeqGAN is the first work extending GANs to generate sequences of discrete tokens. It solves the common problems of traditional GAN in dealing with discrete data such as sequences, that is, the generator is difficult to transfer gradient updates effectively, and the discriminator is difficult to evaluate incomplete sequences. SeqGAN combines the GAN and Policy Gradient algorithm of reinforcement learning to guide the training of the generative model through the discriminative model. SeqGAN uses a Long Short-Term Memory (LSTM) as the generator, a CNN with a highway structure as the discriminator, and a well-trained oracle with the same architecture as the generator to generate samples as the ground truth. The discriminator updates parameters by distinguishing between real samples and generated ones from the generator in the d-step (the step of training the discriminator), which belongs to a binary classification task. The generator uses the Monte Carlo (MC) search reward performed by the discriminator in combination with the policy gradient method to update its parameters in the g-step (the step of training the generator).

\section{Approach}
\label{sec:approach}
We propose a system called \tool that learns to synthesize brand new GUI designs for designers by modeling GUI component subtree sequences and style compatibility. 
The approach overview can be seen in Fig~\ref{fig:approach}
Based on subtrees automatically segmented from the original GUIs in Section~\ref{sec:task}, we first convert all them into embedding by modeling their style in Section~\ref{sec:styleEmbedding}.
During the training process, the generator randomly generates a sequence with the given length and the discriminator acts as the environment, in which the reward can be calculated as the $loss_g$ by Monte Carlo tree search (MCTS). 
We get the homogeneity value of the generated result as $loss_c$ in Section~\ref{sec:styleCompatibility}. 
By measuring the distance between the generated result and the original GUI design, the model captures the structural information in Section~\ref{sec:structure} with $loss_s$ calculated by the minimum edit distance. 
By integrating all the loss functions above in Section~\ref{sec:loss}, the parameters of the generator are updated with the back-propagation algorithm. 

\subsection{Style Embedding of Subtree}
\label{sec:styleEmbedding}
Since we are feeding the model with sub-images showing GUI component subtrees, we first convert all of them to an embedding. 
Similar to natural-language sentence, the overall GUI component layout tree can be regarded as a sentence, and the subtrees obtained from its metadata decomposition is the constituent words of this sentence. We serialize the subtrees by depth-first traversal and map them into embedding space to get their vector features as the input of our \tool. \revised{Thus, we apply a deep learning network to get the feature vector and style embedding of the subtree sequences}.

To transform the image from pixel level to vector level, we adopt a siamese network~\cite{bromley1994signature, koch2015siamese} to model the GUI design with a dual-channel CNN structure, which maps a GUI into GUI vector space. We apply a pair of GUI images ($g1$,$g2$) as the input and the goal of the siamese network is to distinguish whether the two images are from the same app. 
According to our observation, the GUIs from one app is more similar in design style than GUIs from different apps.
Therefore, we set up the learning function to discriminate if two input design images are from one app or not to make the input embedding more meaningful, i.e., representing the design style. 
The CNN in the siamese network takes one of the GUI screenshot pair as the input. Then the convolution operation is executed with various filters ($m \times m$ matrix) to extract the features in the GUIs. A Relu activation and a max-pooling layer follow the convolution operation, which can be considered as a convolutional block and they are stacked repeatedly. In the end, the output of the last convolutional block, which represents the embedding in the vector space of the GUIs, is flattened into an FC(fully-connected) layer. 

The goal of the trained CNN in the siamese network is to convert the GUI screenshot image $g$ to N-dimensional vector $V_g$. This non-linear transformation function $f$ can be expressed as $V_g = f(g;\theta)$, where $\theta$ represents the trainable parameters in the network which can be updated by the back-propagation algorithm in the training process. The weighted $L1$ distance is applied to measure the two feature vectors $V_{g1}$ and $V_{g2}$ from the two channels and then fed into a $sigmoid$ activation function to calculate the predictive result. Since this task can be formulated as a binary classification problem i.e., two input screenshots from the same app or not, we adopt the binary cross entropy loss function:

\begin{equation}
Loss(x,y) = -\sum_{i}(x_{i}\log(y_{i})+(1-x_{i})\log(1-y_{i}))
\label{eq:crossEntropy}
\end{equation}where $x$ is the probability output of the network and $y$ is the target (0 or 1).

The CNNs of the different channels use the same weights and learn the ability to obtain the most representative information features in GUIs, which is used to quantitatively compare the appearance design style similarity between GUI images. The pixel information of the intercepted layout subtrees from the original GUI is fed into the trained CNN and thus we acquire their design embeddings. 

\subsection{Modeling Subtree Compatibility}
\label{sec:styleCompatibility}
We apply the CNN (in the Siamese network) trained in the previous section to help evaluate the aesthetic identity of the generated samples.
In each g-step, when the generator generates a complete sequence $S_{1:T}$, which is composed of $T$ subtrees from different apps (subtree repository) spliced in order.
According to the metadata of their GUIs, we can acquire the coordinates of each subtree and intercept their images from their original GUI images.
Then we input them into the CNN from the Siamese network trained before and output their embeddings. Using these embeddings, we apply the homogeneity (HOM) to evaluate the aesthetic compatibility of subtrees in the sequence.

Homogeneity (HOM) is the proportion of clusters containing only members of a single class \revised{(the class here represents the app, and when the subtrees all come from the same app, they get the highest harmony)} by

\begin{equation}
h = 1- \frac{H(G|C)}
{H(G)}
\label{eq:hom}
\end{equation}

where $H(G|C)$ is the conditional entropy of a class with a given cluster assignment and $H(G|C) = -\sum_{g=1}^{\left | G \right |}\sum_{c=1}^{\left | C \right |}\frac{n_g,c}{n}\log(\frac{n_g,c}{n_c})$. $H(G)$ represents the entropy of $G$ and $H(G) = -\sum_{g=1}^{\left | G \right |}\frac{n_g}{n}\log(\frac{n_g}{n})$. $n$ is the total number of samples, $n_g$ and $n_c$ belong to class $G$ and class $C$ respectively, and $n_{g,c}$ is the number of samples divided from class $G$ to class $C$. 

We expect that the generator can keep learning to make the generated results have higher homogeneity scores, which represents better coordination and compatibility. Therefore, we integrate the homogeneity score of the generated result into the training of the generator after the discriminator feeds back the reward value to a complete sequence from the generator. Then we calculate the style loss as

\begin{equation}
Loss_c = \left\{\begin{matrix}
\exp{(-h)}, if~c > 1\\ 
0, if~c = 1
\end{matrix}\right.
\end{equation}
where $c$ is the number of the app where the subtrees come from. If the subtrees of a sample are all from the same app, $Loss_c$ becomes zero.

\subsection{Modeling Subtree Structure}
\label{sec:structure}
In addition to style information, another factor to be considered is the structure information corresponding to the generated sequence. The layouts of each actual GUI have certain composition rules, which make a GUI not only more logical in appearance but also practical function. We hope that when generating new subtree combination sequences, the generator can also follow the composition conditions of GUI to a certain extent so that these synthetic sequences not only stay diversity but also meet the structural characteristics of the real GUIs. 
For this purpose, we use the structure strings of the subtrees from their metadata instead of the GUI wireframe images~\cite{deka2017rico}~\cite{ 2020Wireframe} to represent their structures as there is explicit order among different GUI components. 
The minimum edit distance (MED) is introduced to quantify the structural similarity between two GUIs. 
The MED can be used to evaluate the structural similarity \revised{between the generated samples and the real world data}. By reducing the structural distance, we can optimize the generator, so that it learns the reasonable structure combination and order from the real GUIs, which can be expressed as 

\begin{equation}
Loss_s = \left\{\begin{matrix}
\max(i,j), if\min(i,j)=0,\\ 
\min
\left\{\begin{matrix}
lev_{S_r,S_g}(i-1,j)+1\\ 
lev_{S_r,S_g}(i,j-1)+1\\ 
lev_{S_r,S_g}(i-1,j-1)+1
\end{matrix}\right.
\end{matrix}\right. otherwise.
\label{eq:TreeLoss}
\end{equation}

where $S_r$ and $S_g$ represent the subtree structure strings of real and generated samples, $lev_{S_r,S_g}(i,j)$ is the distance between the first $i$ characters in $S_r$ and the first $j$ characters in $S_g$. \revised{Each character represents a GUI component such as ListView or FrameLayout.}

\subsection{Multi-Loss Fusion}
\label{sec:loss}
There is a large span in the numerical region of the three losses(the feedback loss from the discriminator, $loss_g$, $loss_c$, and $loss_s$), so that we need to normalize them for subsequent calculation.
By adding the trainable noise parameters~\cite{kendall2018multi}, the three loss values can be balanced to the same scale and we express the final fusion loss function as 
\begin{equation}
Loss_{mul} = \lambda_1 Loss_g+\lambda_2 Loss_c+\lambda_3 Loss_s
\label{eq:multi-loss}
\end{equation}

In g-step, we update the parameters of the generator by minimizing $Loss_{mul}$ and we apply $Adam$ update algorithm\cite{kingma2014adam} instead of stochastic gradient descent (SGD) for faster convergence.

\begin{equation}
\arg\min_{G}\ Loss_{mul}
\label{eq:bp}
\end{equation}

\begin{figure}[h]
    \begin{minipage}{1\linewidth}
        \centerline{\includegraphics[width=0.95\linewidth]{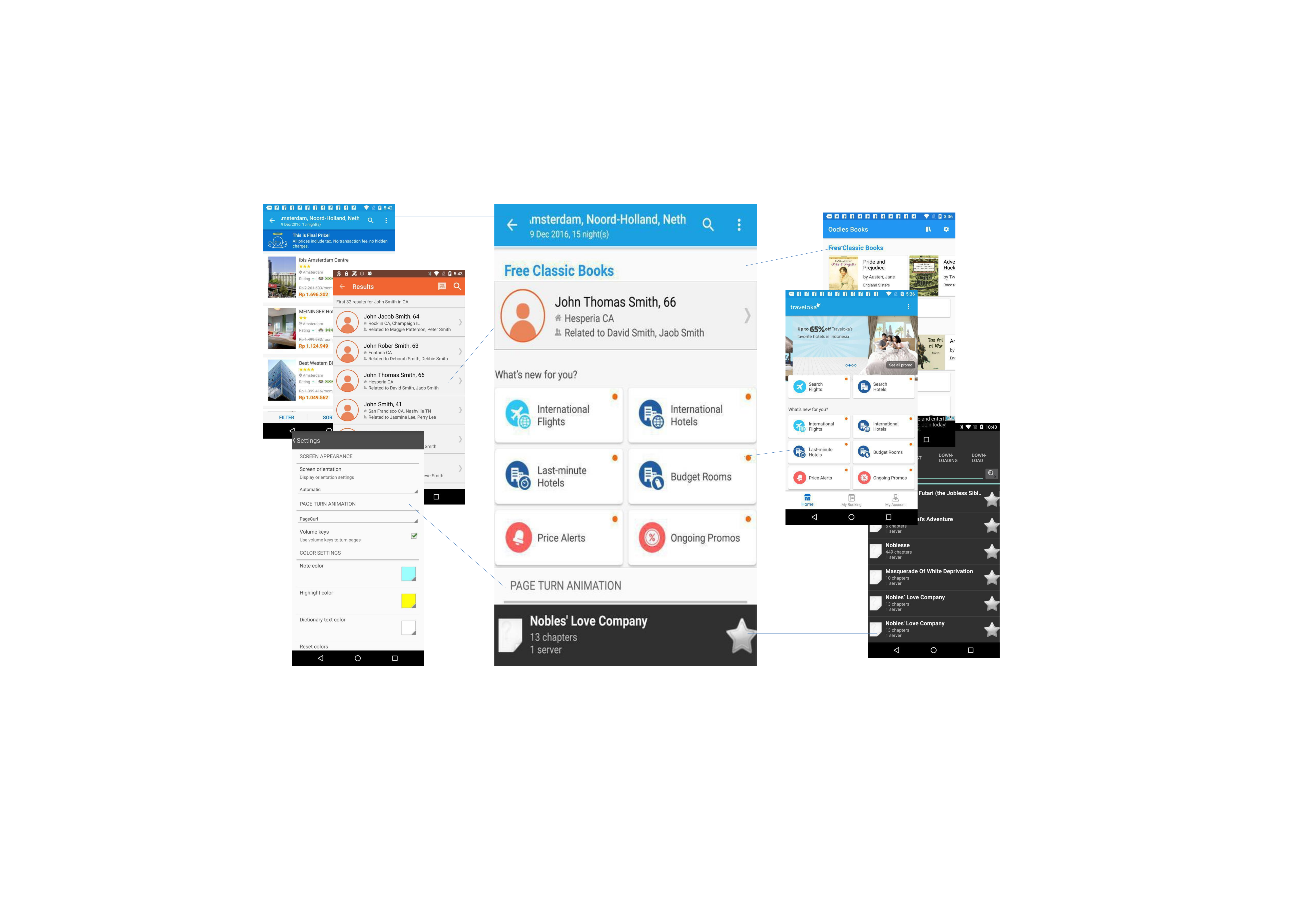}}
        \centerline{(a)}
    \end{minipage}%
    \hfill
    \\
    \begin{minipage}{1\linewidth}
        \centerline{\includegraphics[width=0.95\linewidth]{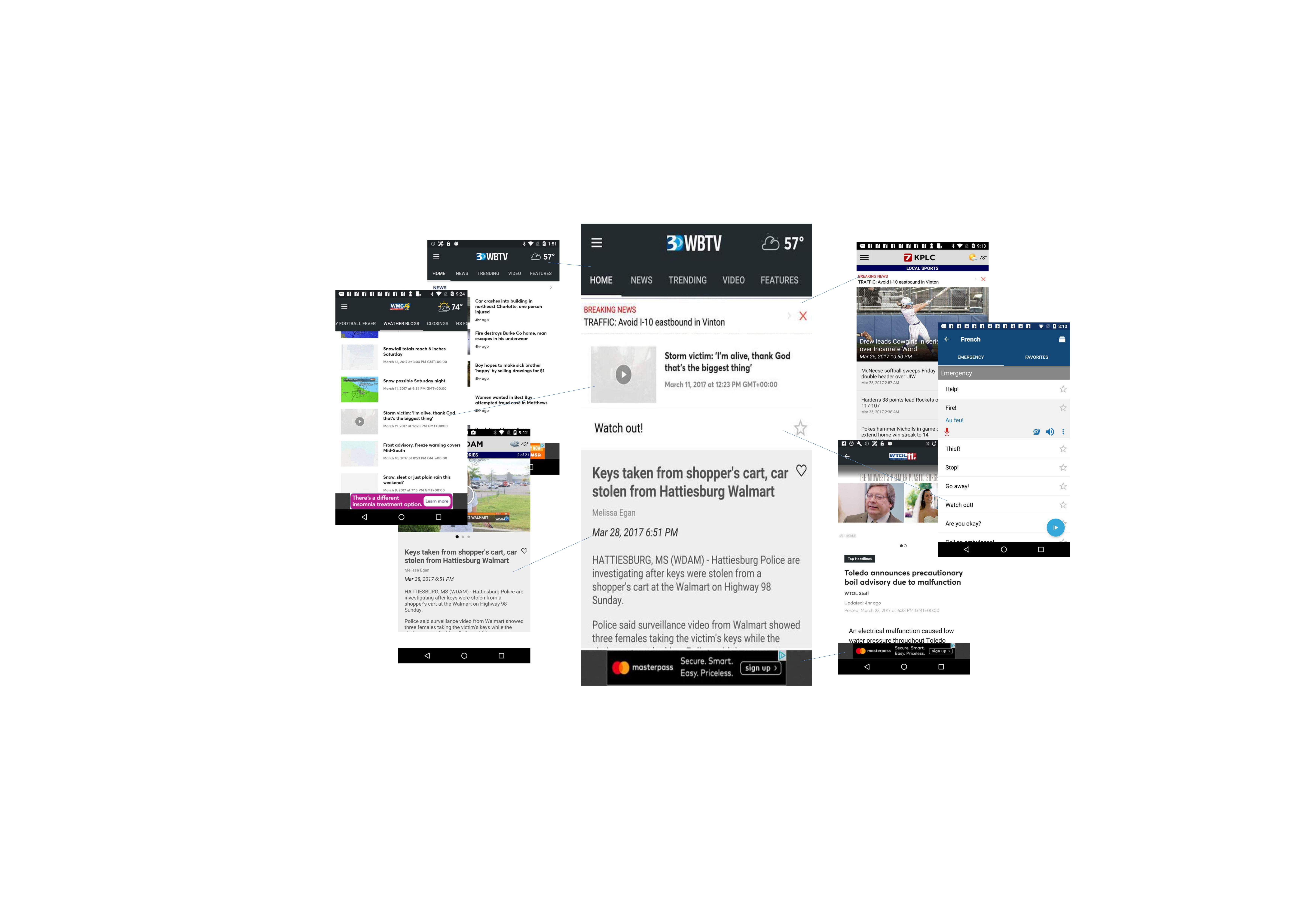}}
        \centerline{(b)}
    \end{minipage}%
    \hfill
    \caption{Two example GUIs generated by \tool with components from corresponding original GUIs.}
    \label{fig:TwoSamples}
\end{figure}

The challenge of the model is to generate a new GUI design with a reasonable structure and compatible style. We don't want the model to only learn to generate sequences similar to the real samples. It promotes \tool to generate new GUI designs with authenticity and diversity by fusing the structure and style information simultaneously to the original sequence features. As shown in Fig~\ref{fig:TwoSamples}, two samples generated by the \tool are reconstructed by the pieces from the real GUI. 

\section{Implementation}
The data in our experiment comes from the Rico open-source online data collection~\cite{deka2017rico}, and a part of GUIs for this article is retained through manual screening. Our model mainly consists of a SeqGAN (includes an LSTM and a CNN) and a Siamese network. All networks are implemented on the PyTorch platform and trained on a GPU.

\subsection{Dataset Construction}
Our data comes from Rico~\cite{deka2017rico}, an open-source mobile app dataset for building data-driven design applications. Rico is the largest repository of mobile app designs to date, supporting design search, UI layout generation, UI code generation, etc. Rico was built by mining Android apps at runtime via human-powered and programmatic exploration. 13 workers spent 2,450 hours using the downloaded apps from the Google Play Store on the platform over five months, producing 10,811 user interaction traces. Rico contains design and interaction data for 72219 UIs from 9772 apps, spanning 27 categories.

\begin{figure}[h]\scriptsize
    \begin{minipage}{0.25\linewidth}
        \centerline{\includegraphics[width=2.2cm,height=4.4cm]{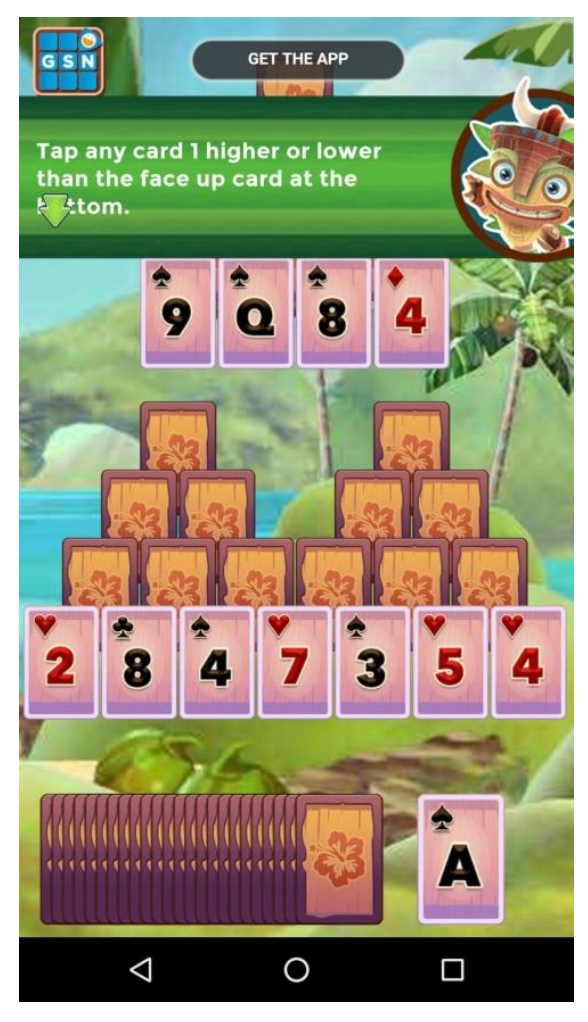}}
        \centerline{(a)}
    \end{minipage}%
    \hfill
    \begin{minipage}{0.25\linewidth}
        \centerline{\includegraphics[width=2.2cm,height=4.4cm]{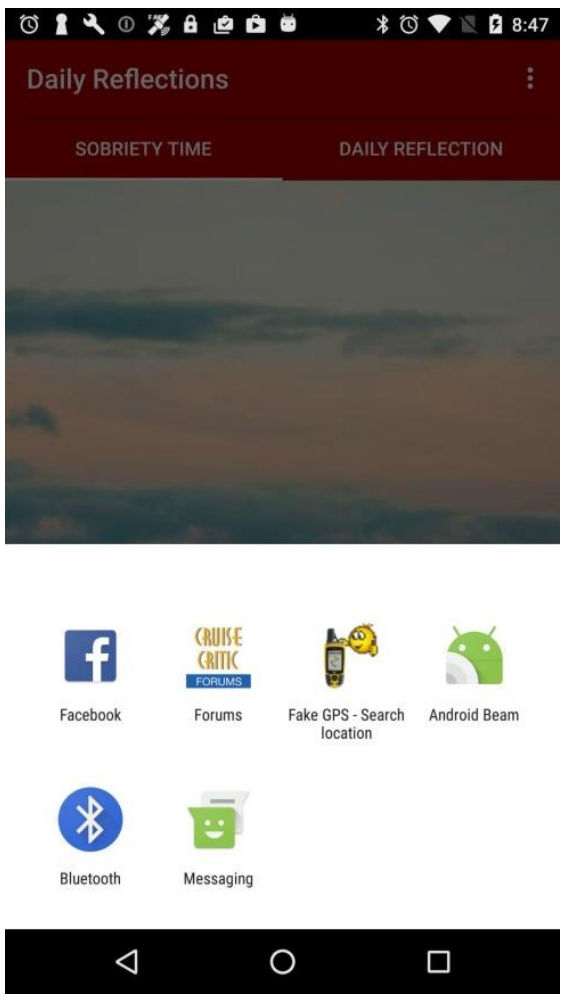}}
        \centerline{(b)}
    \end{minipage}%
    \hfill
    \begin{minipage}{0.25\linewidth}
        \centerline{\includegraphics[width=2.2cm,height=4.4cm]{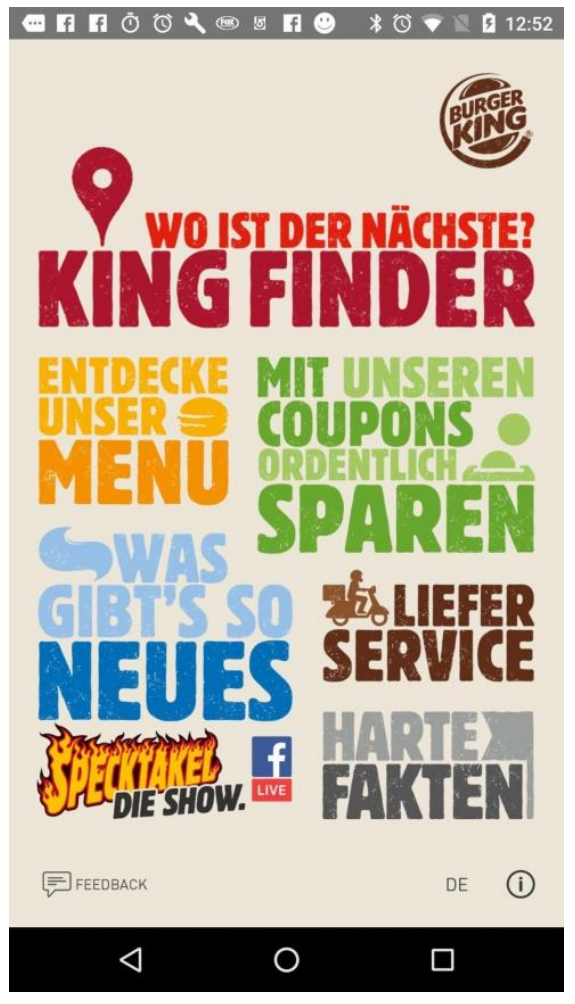}}
        \centerline{(c)}
    \end{minipage}%
    \hfill
    \begin{minipage}{0.25\linewidth}
        \centerline{\includegraphics[width=2.2cm,height=4.4cm]{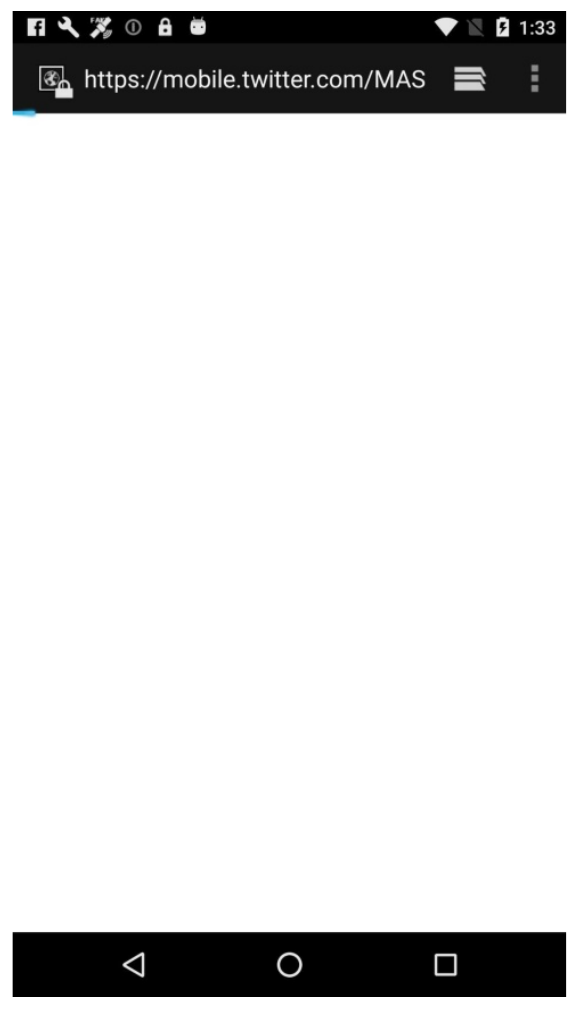}}
        \centerline{(d)}
    \end{minipage}%
    \caption{Examples of GUIs we removed from our dataset: The GUI from the game app (a), pop-up window (b), large picture (c), waiting page (d).}
    \label{fig:RemovedExamples}
\end{figure}

\revised{Based on our observation, not all GUIs from Rico datasets~\cite{deka2017rico} can be used in this study, so we remove some of them. First, we remove the GUIs of game apps as game app GUIs is specifically generated by game engine which is different from other general apps. Second, we manually remove some low-quality GUIs including large pop-up windows, advertisements, or posters occupying the whole screen size, webpage, loading page with a progress bar, and real scenes in camera. Some
examples can be seen in Fig~\ref{fig:RemovedExamples} and we release all of our datasets at our online gallery\footnote{ \url{https://github.com/GUIDesignResearch/GUIGAN\#dataset-construction}}.}

% removed

\subsection{Model Implementation}
\revised{As we take the subtree of the existing GUIs as the basic unit for composing a new GUI design in our paper. A sample of the real-world data is the combination of the subtrees from one single real-world GUI in order, with both the screenshots and structure information based on the metafile. It is mainly used as the training data for our GAN based approach. We first split those GUIs collected from real-world Android apps in Google Play into subtrees following procedures in Section~\ref{sec:task}. We then train a Siamese network with this data to get subtree style embedding in Section~\ref{sec:styleEmbedding}. The subtree embedding is input to the generator of the \tool for modelling both the style loss ($Loss_c$) and structure loss ($Loss_s$) for generating new subtree sequences, which can be used for composing new GUI designs. The generated subtree sequences are then input to the discriminator within the SeqGAN for training the discriminator to discriminate the real-world GUI and generated GUI. After the adversarial training, given random noise or pre-built GUI components, the \tool can generate a new GUI by composing several subtrees from real-world GUIs.}

The structure of the generator and the discriminator in SeqGAN is preserved in \tool which is implemented in PyTorch. 
We store the start and end subtrees of each real-world GUI in the start list and end list respectively. The LSTM randomly takes a subtree in the start list as the initial matrix, not the zero matrices in SeqGAN.
Thus, the generator generates a sequence of length $T$ (the default sequence length is 30 because the GUI subtree length in real-world data is mostly within 30).
From the beginning of the first subtree in the start list, if the total height of the later spliced subtree and all previous subtrees exceeds the rated height, the subsequent splicing will be stopped. If the subtree in the end list is selected, the splicing will be stopped directly. 

The Long Short-Term Memory (LSTM) is used as the generative network. The vector dimension is selected to be 32 and the hidden layer feature dimension is selected to be 32. Like the discriminative model in SeqGAN, we also use a CNN network that joins the highway architecture. The batch size is 32 and the learning rate is set to 0.05. 

The siamese network is a two-channel CNNs with shared weights. The positive example is a pair of subtrees from the same app with a label of 1, and the negative one from two different apps with a label of 0. The subtree image is resized to $512 \times 256$ by the Nearest Neighbor algorithm as the input. There are 4 \textit{Conv} $\rightarrow$ \textit{Pool} layers blocks in the CNN structure. The first \textit{Conv} layer use 64 filters, and each subsequent \textit{Conv} layer doubles the number of filers. We set the filter size as $10 \times 10$, $7 \times 7$, $4 \times 4$, and $4 \times 4$ in different CNN layers, and the stride as  $1 \times 1$ for convolutional layers. We apply the pooling units of size  $2 \times 2$ applied with a stride 2. We train the Siamese network with 50 epochs for about three hours.

\section{Automated Evaluation}
\label{sec:auto-evaluation}
In this section, we prepare GUI images collected according to the category and development company of the app as experimental data based on the statistics and test the performance of the proposed model on GUI generation. We use the real-world data as the ground truth and introduce WGAN-GP, FaceOff, and two variations of our \tool as the baseline methods to compare the two metrics of FID and 1-NNA.

\subsection{Experimental Dataset}
When preparing the experiment dataset, we consider two specific usage scenarios when developing app GUIs.
First, most designers and developers have a clear goal for developing the GUI of an app in a specific category such as finance, education, news, etc.
Since each kind of app category owns its characteristics, we try to test our model's capability in capturing that characteristic by preparing a separated dataset for the five most frequent app categories in Rico dataset~\cite{deka2017rico} including News \& Magazines, Books \& Reference, Shopping, Communication, and Travel \& Local as shown in  Table~\ref{tab:dataset}.

Second, we notice that designers and developers often refer to the GUI design style developed by big companies when developing their own GUI.
Therefore, another experiment is to generate GUIs by learning GUI designs from a specific company.
Based on their metadata, we prepare three kinds of GUIs from three big companies with most apps in our dataset i.e., Google, YinzCam, and Raycom as shown in Table~\ref{tab:dataset}.

\renewcommand{\arraystretch}{1.5} %ÃÂ¦ÃÂÃÂ§ÃÂ¥ÃÂÃÂ¶ÃÂ¨ÃÂ¡ÃÂÃÂ©ÃÂ«ÃÂ  
\begin{table}[h]
\caption{GUI dataset by category or company}
\centering
\fontsize{9}{8}\selectfont
\setlength{\tabcolsep}{3.5mm}{
\begin{tabular}{lll}
\hline
Category or Company  & \#App & \#GUI \\
\hline
News \& Magazines&108&467\cr
Books \& Reference&110&394\cr  
Shopping&98&460\cr 
Communication&107&357\cr
Travel \& Local&85&407\cr \hline
Google&28&85\cr
Yinzcam&30&136\cr  
Raycom&27&113\cr
\hline
\end{tabular}}
\label{tab:dataset}
\end{table}

\subsection{Evaluation Metrics}
In order to quantify and measure the similarity between the real data distribution $P_r$ and the generated sample distribution $P_g$, we introduce Fr\'echet Inception distance (FID)~\cite{heusel2017gans} and 1-Nearest Neighbor Accuracy (1-NNA)~\cite{lopezpaz2017revisiting} as the evaluation metrics. FID measures the diversity and quality of generated images relative to real images, and 1-NNA is used to analyze the distribution differences between the two sample sets.

\textbf{Fr\'echet Inception distance (FID)} is a widely-used metric~\cite{salimans2016improved} recently introduced \revised{for measuring the quality and diversity of generated images, especially by GAN}. FID is calculated according to Inception Score (IS), comparing the statistics of negative samples to positive samples, using the Fr\'echet distance between two multivariate Gaussians:

\begin{equation}
    {\rm FID}(P_r,P_g) = 
    \left \| \mu_r - \mu_g \right \| + Tr(C_r+C_g-2(C_rC_g)^{1/2})
\end{equation}%

where $\mu_r$ and $\mu_g$ are the mean values of the 2048-dimensional activations of the Inception-v3~\cite{szegedy2016rethinking}(By asymmetrically decomposing the convolution operation, the depth and width of the network pretrained on the ImageNet are increased simultaneously, and the 1 $\times$ 1 network is widely used to reduce the dimension) pool3 layer for real and generated samples $r$ and $g$ respectively, while $C_r$ and $C_g$ are the covariances. A lower FID represents that the two distributions are closer, which means that the quality and diversity of the generated images are higher. In our experiments, we input the same number of real and generated image collections into the Inception-v3 network to get the FID score.

\textbf{1-Nearest Neighbor Accuracy} is used in two-sample tests to assess whether two distributions are identical. A well trained 1-nearest neighbor classifier is applied. The better the performance of the generated model, the more difficult it is for the 1-NN classifier to distinguish the true from the false. Therefore, the best recognition rate is 50\%, the worst is 100\%. But if the recognition rate is lower than 50\%, it means that the model may be overfitting.
%\chen{Is this number smaller, the better? If so, please mention it in the paper.}
This metric is widely used for evaluating the quality of generated images~\cite{yang2019pointflow, zhang2019stylistic, Xu2018An}.

\renewcommand{\arraystretch}{1.5} %ÃÂ¦ÃÂÃÂ§ÃÂ¥ÃÂÃÂ¶ÃÂ¨ÃÂ¡ÃÂÃÂ©ÃÂ«ÃÂ  
\begin{table*}[h]
  \centering  
  \fontsize{9}{8}\selectfont  
  \begin{threeparttable}  
  \caption{Performance by different app categories}
    \begin{tabular}{ccccccccccc} 
    \toprule  
    \multirow{2}{*}{Category}& 
    \multicolumn{2}{c}{WGAN-GP}&\multicolumn{2}{c}{FaceOff}&\multicolumn{2}{c}{\tool}&\multicolumn{2}{c}{\tool-Style}&\multicolumn{2}{c}{\tool-Structure}\cr  
    \cmidrule(lr){2-3} \cmidrule(lr){4-5}  \cmidrule(lr){6-7} \cmidrule(lr){8-9} \cmidrule(lr){10-11}
    &FID&1-NNA&FID&1-NNA&FID&1-NNA&FID&1-NNA&FID&1-NNA\cr  
    \midrule  
    News \& Magazines&0.181&0.999&0.145&0.987&\textbf{0.110}&\textbf{0.857}&0.139&0.876&\textbf{0.110}&0.900\cr
    Books \& Reference&0.161&1.000&0.106&0.972&\textbf{0.084}&\textbf{0.871}&0.107&0.923&0.101&0.906\cr  
    Shopping&0.172&0.998&0.120&0.972&0.068&\textbf{0.829}&0.052&\textbf{0.829}&\textbf{0.048}&0.868\cr 
    Communication&0.150&0.995&0.134&0.993&\textbf{0.060}&\textbf{0.915}&0.065&\textbf{0.915}&0.089&0.929\cr
    Travel \& Local&0.139&0.999&0.060&0.974&\textbf{0.054}&\textbf{0.873}&0.067&0.897&0.088&0.880\cr
    \cmidrule(lr){1-11}
    Average&0.161&0.998&0.113&0.980&\textbf{0.075}&\textbf{0.869}&0.086&0.888&0.087&0.897\cr
    \bottomrule 
    \end{tabular}
    \label{tab:MetricsOfCategories}
    \end{threeparttable} 
\end{table*} 

\renewcommand{\arraystretch}{1.5} %ÃÂ¦ÃÂÃÂ§ÃÂ¥ÃÂÃÂ¶ÃÂ¨ÃÂ¡ÃÂÃÂ©ÃÂ«ÃÂ  
\begin{table*}[h]
  \centering  
  \fontsize{9}{8}\selectfont  
  \begin{threeparttable}  
  \caption{Performance by different app development companies}
    \begin{tabular}{ccccccccccc}
    \toprule  
    \multirow{2}{*}{company}& 
    \multicolumn{2}{c}{WGAN-GP}&\multicolumn{2}{c}{FaceOff}&\multicolumn{2}{c}{\tool}&\multicolumn{2}{c}{\tool-Style}&\multicolumn{2}{c}{\tool-Structure}\cr  
    \cmidrule(lr){2-3} \cmidrule(lr){4-5}  \cmidrule(lr){6-7} \cmidrule(lr){8-9} \cmidrule(lr){10-11}
    &FID&1-NNA&FID&1-NNA&FID&1-NNA&FID&1-NNA&FID&1-NNA\cr  
    \midrule  
    Google&0.181&0.999&0.125&0.945&0.131&0.844&0.122&\textbf{0.837}&\textbf{0.087}&0.859\cr
    Yinzcam&0.172&0.980&0.146&0.964&\textbf{0.074}&\textbf{0.806}&0.084&0.828&0.094&0.828\cr
    Raycom&0.244&1.000&0.101&0.968&0.063&\textbf{0.819}&\textbf{0.049}&0.846&0.075&0.840\cr
    \cmidrule(lr){1-11}
    Average&0.199&0.990&0.124&0.959&0.089&\textbf{0.823}&\textbf{0.085}&0.837&\textbf{0.085}&0.842\cr
    \bottomrule  
    \end{tabular} 
    \label{tab:MetricsOfcompanies}
    \end{threeparttable} 
\end{table*}

\subsection{Baseline Models}
According to our query of relevant information, there are very little researches on generating GUI designs. 
But there are some related works about image generation which is roughly similar to our task, so we use two different kinds of methods as baselines with one from the image generation and the other one using template search.

The first baseline is \textbf{WGAN-GP}~\cite{GulrajaniImproved}, which is proposed on the basis of WGAN~\cite{arjovsky2017wasserstein}. WGAN introduces Wasserstein distance and optimizes the implementation process of the algorithm to solve the problem of gradient vanishing in the training process of traditional GAN. WGAN-GP introduces a gradient penalty in WGAN, which accelerates the convergence of the model and has a more stable training process. 

The second baseline is \textbf{FaceOff}~\cite{zheng2019faceoff} which parses the DOM tree of a raw input website created by a user through measuring the distance of the trees and uses a CNN to learn style compatibility to find a similarly well-designed web GUI. Although FaceOff is used to generate a new web GUI, in this article we modify its raw input to the real world data GUI structure of the mobile application as a query and then sort the retrieved results according to the homogeneity score of their subtree combination. 

Apart from the two baselines mentioned above, we also get some derivation baselines from our model by changing the muti-loss in the generator.
One (\textbf{\tool-style}) is, using only the $Loss_c$ correction, and the other one (\textbf{\tool-structure}) with only the $Loss_s$ correction within which the generator focuses on either design style or structure characteristics. 
In this way, we can compare and observe within our model how these modifications affect the generated results in metric.

\subsection{Results}
\begin{figure}\scriptsize
    \begin{minipage}{0.25\linewidth}
        \centerline{\includegraphics[width=2.2cm,height=4.4cm]{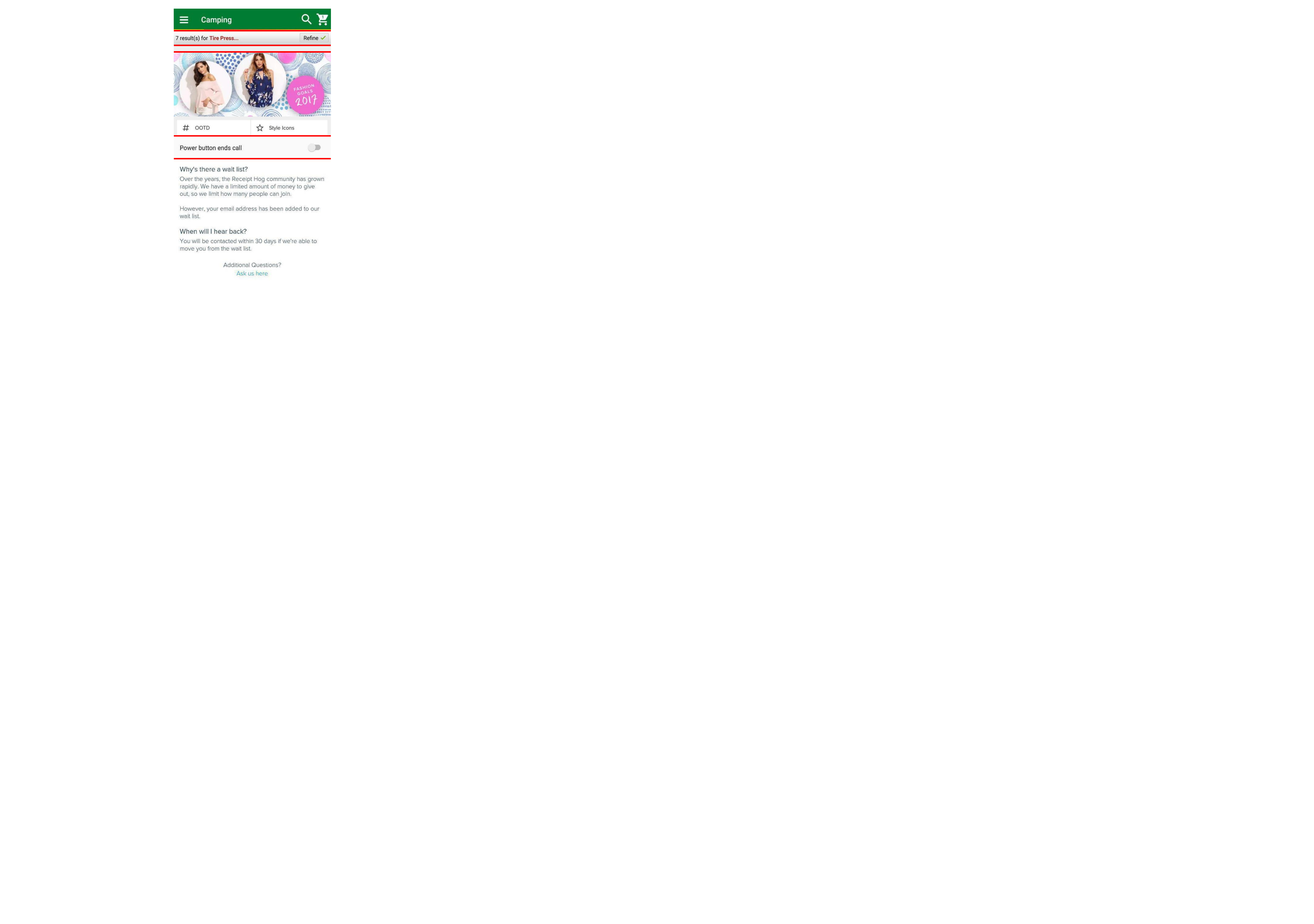}}
        \centerline{(a)}
    \end{minipage}%
    \hfill
    \begin{minipage}{0.25\linewidth}
        \centerline{\includegraphics[width=2.2cm,height=4.4cm]{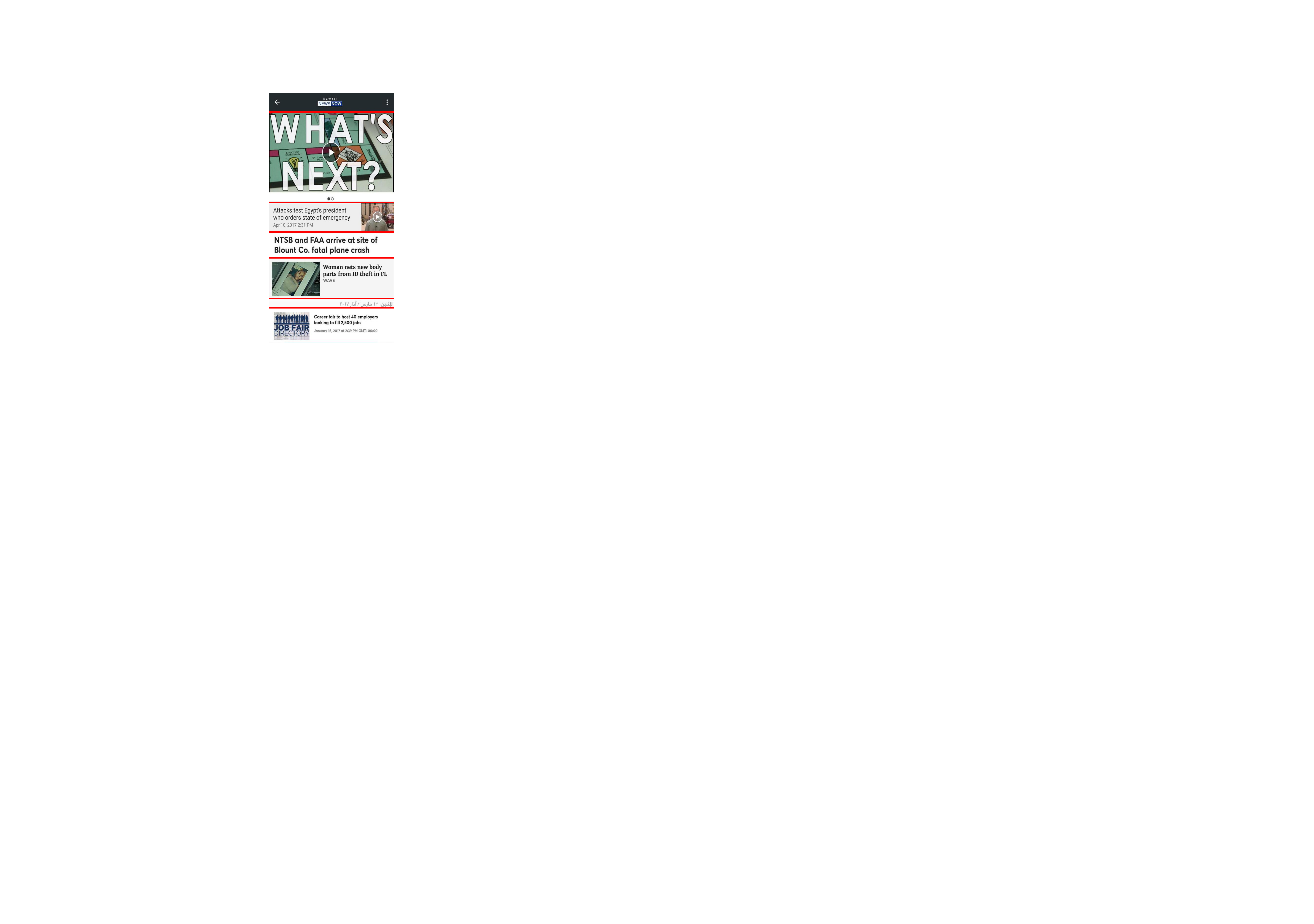}}
        \centerline{(b)}
    \end{minipage}%
    \hfill
    \begin{minipage}{0.25\linewidth}
        \centerline{\includegraphics[width=2.2cm,height=4.4cm]{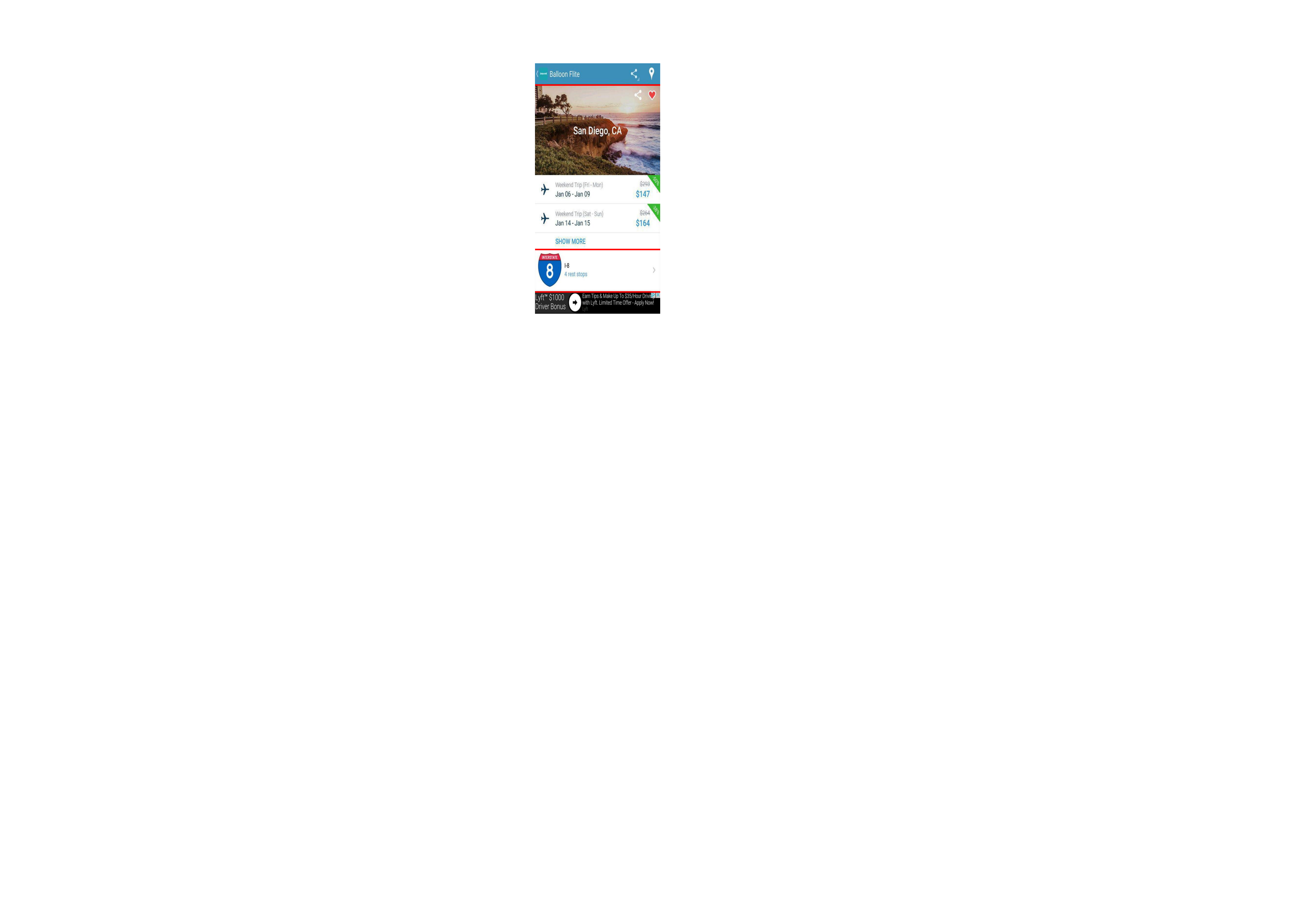}}
        \centerline{(c)}
    \end{minipage}%
    \hfill
    \begin{minipage}{0.25\linewidth}
        \centerline{\includegraphics[width=2.2cm,height=4.4cm]{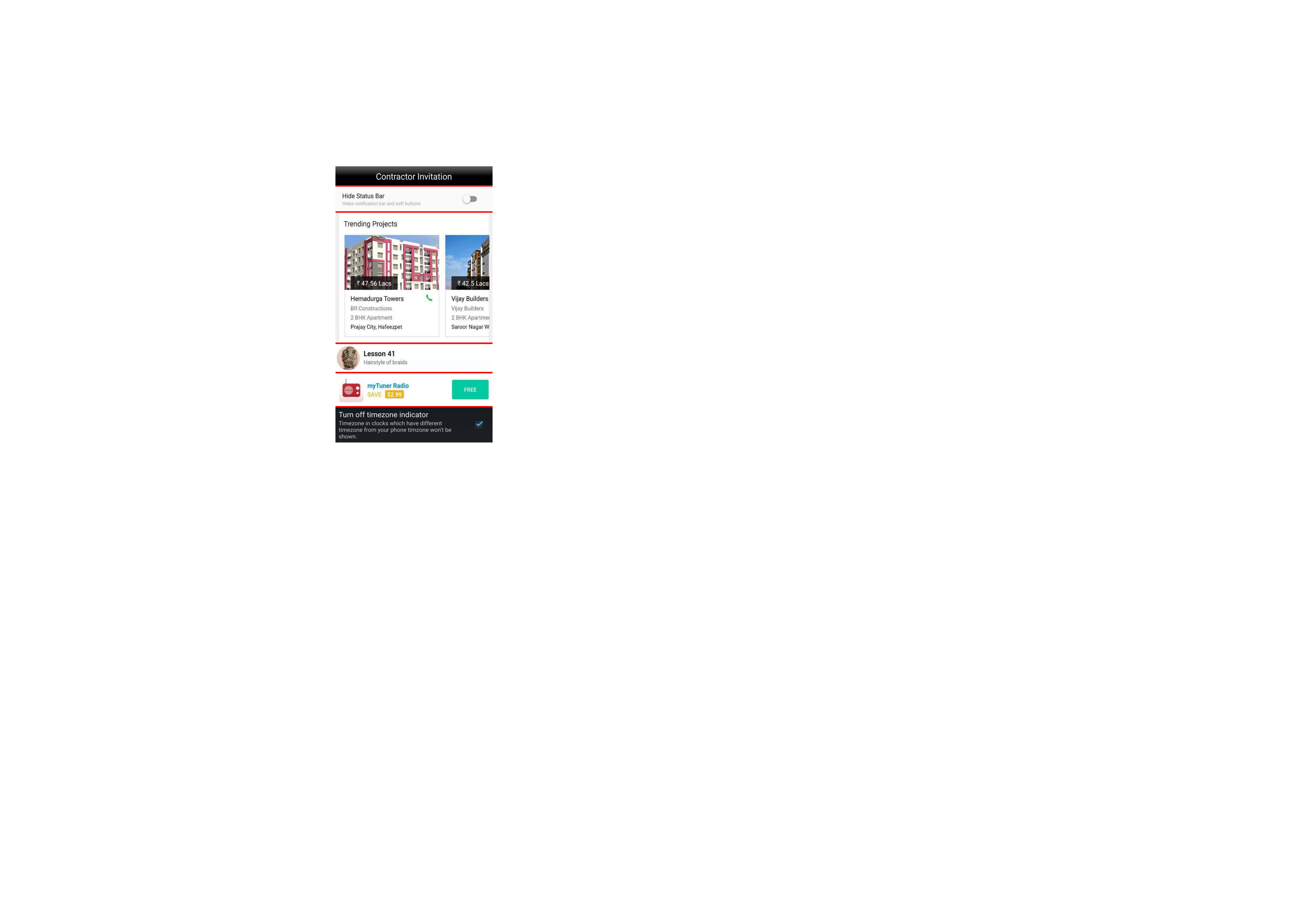}}
        \centerline{(d)}
    \end{minipage}%
    \caption{Examples of the GUIs generated by \tool.}
    \label{fig:SamplesDisplayGUIGAN}
\end{figure}

Table~\ref{tab:MetricsOfCategories} and~\ref{tab:MetricsOfcompanies} show the results of different methods on two metrics in the category and company specific development scenarios. 
The results show that our proposed model achieves the highest scores on FID and 1-NNA under both the GUI development scenario. 
Our model achieves a 33.63\% and 11.33\% boost in FID and 1-NNA than the best baselines in the dataset of category, and 28.23\% and 14.18\% increase in the dataset of the company, respectively. 
Fig~\ref{fig:SamplesDisplayGUIGAN} shows examples of GUI images from \tool, and for ease of observation, we separate different subtrees with thick red lines. 
It can be seen that the GUIs generated from \tool have a comfortable appearance, and a reasonable structure composed of different components.
Meanwhile, it also keeps the overall harmonious design style.
After checking many generated GUIs, we find that both the structure and style of the GUIs are also very diverse which can provide developers or designers with different candidates for their GUI design.
More generated GUIs from \tool can be seen in our online gallery\footnote{ \url{https://github.com/GUIDesignResearch/GUIGAN}}.

% GUIGAN

\begin{figure}\scriptsize
    \begin{minipage}{0.25\linewidth}
        \centerline{\includegraphics[width=2.2cm,height=4.4cm]{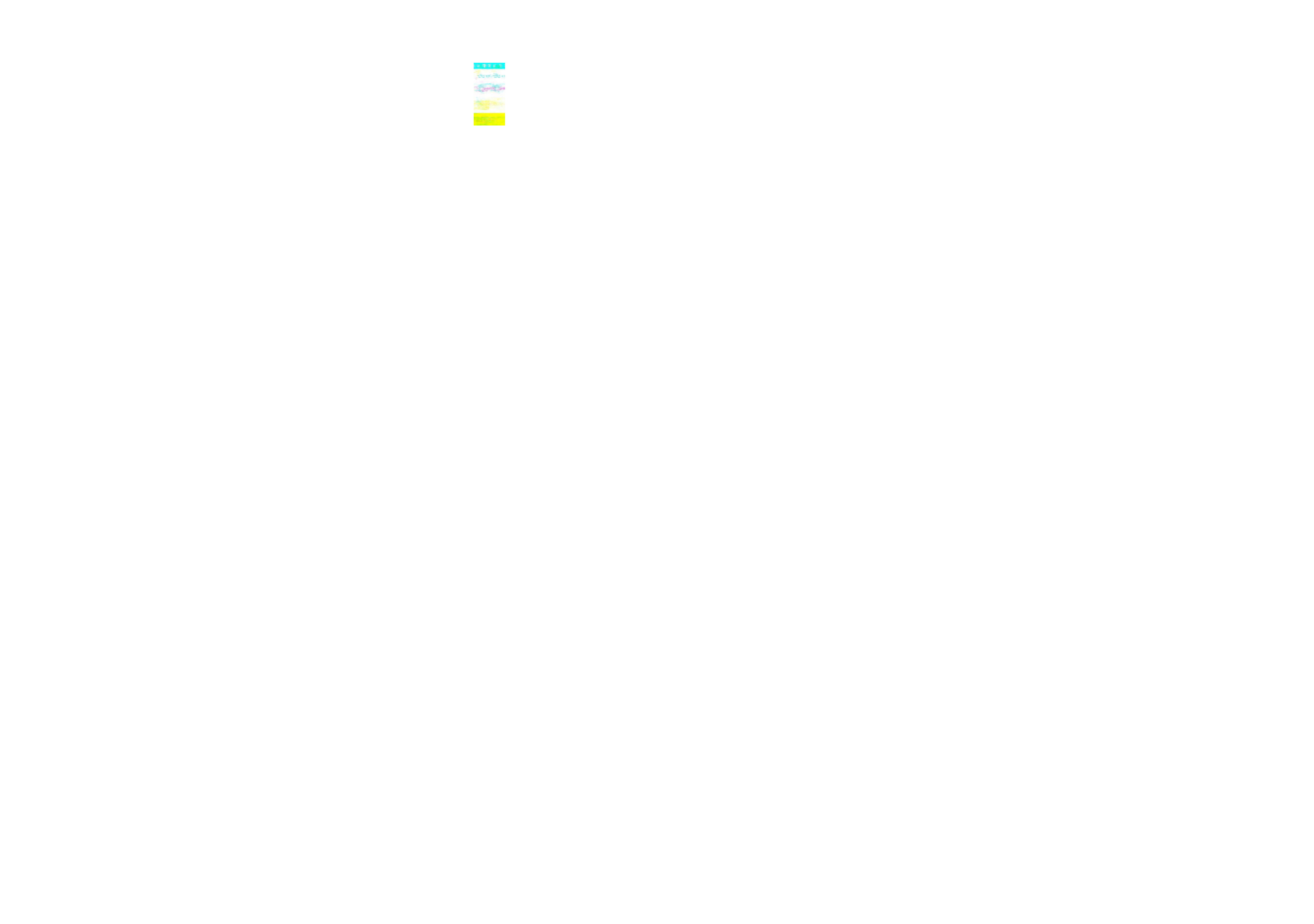}}
        \centerline{(a)}
    \end{minipage}%
    \hfill
    \begin{minipage}{0.25\linewidth}
        \centerline{\includegraphics[width=2.2cm,height=4.4cm]{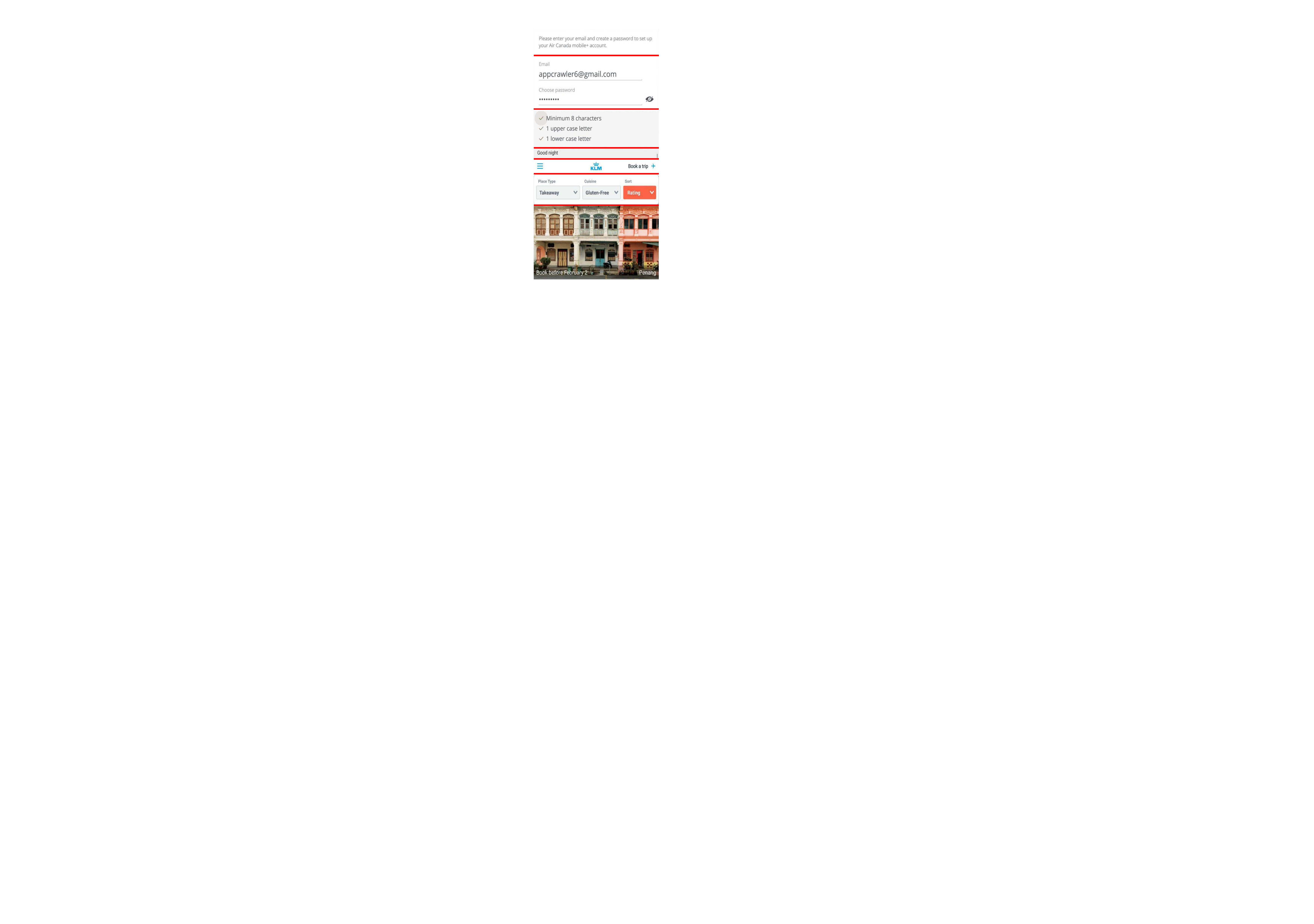}}
        \centerline{(b)}
    \end{minipage}%
    \hfill
    \begin{minipage}{0.25\linewidth}
        \centerline{\includegraphics[width=2.2cm,height=4.4cm]{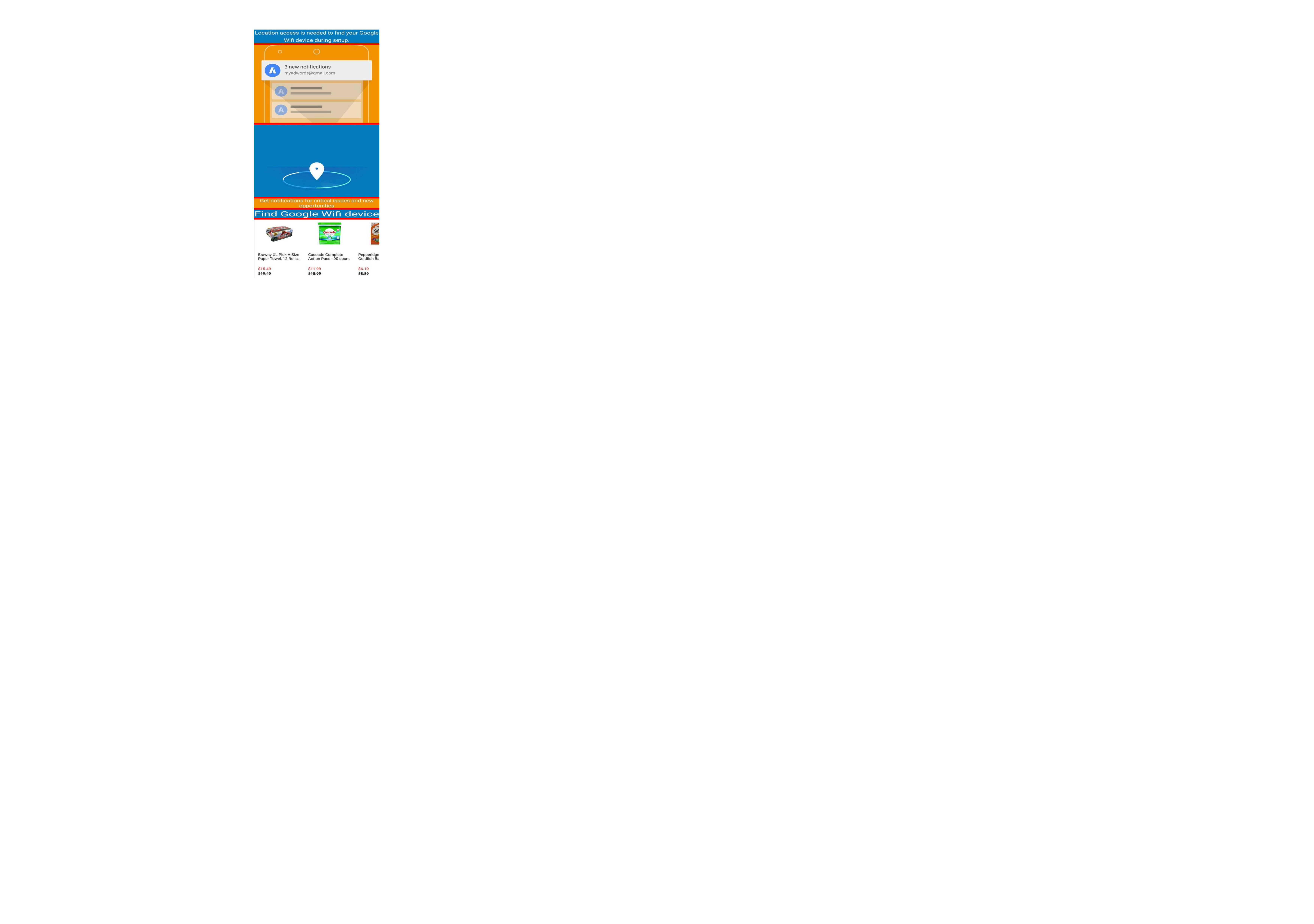}}
        \centerline{(c)}
    \end{minipage}%
    \hfill
    \begin{minipage}{0.25\linewidth}
        \centerline{\includegraphics[width=2.2cm,height=4.4cm]{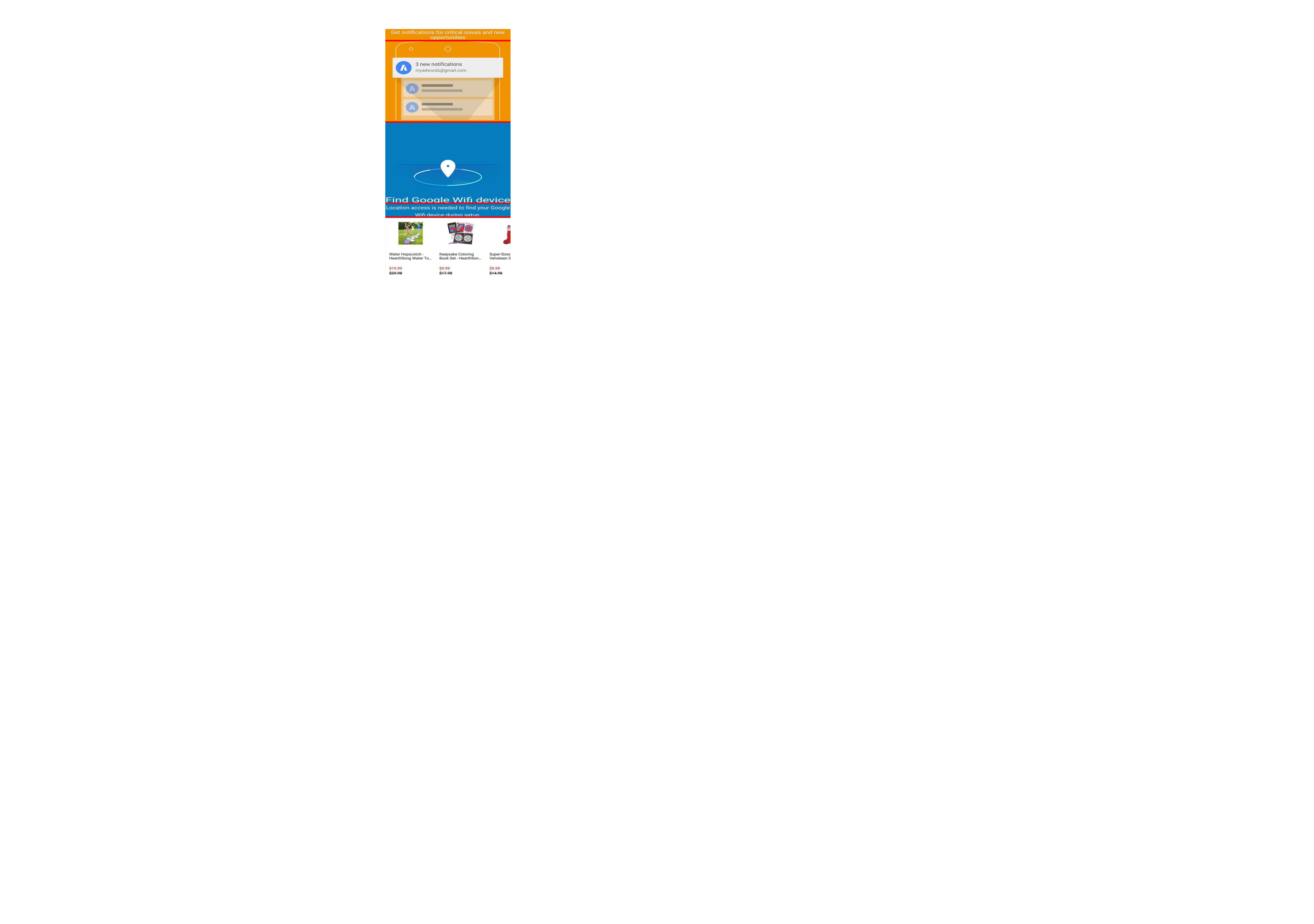}}
        \centerline{(d)}
    \end{minipage}%
    \caption{Generated GUI examples by WGAN-GP (a) and FaceOff (b, c, d).}
    \label{fig:SamplesDisplayBaselines}
\end{figure}

Two baselines including WGAN-GP and FaceOff do not perform well compared with \tool in FID and 1-NNA. 
According to our observation, we can see the overall layout 
 of generated GUIs from WGAN-GP as seen in Fig~\ref{fig:SamplesDisplayBaselines} (a), but very blurred in detail.
That is because WGAN-GP is a pixel-based approach that cannot accurately model the information of component-based GUIs.
Although it is widely used in the natural image, it is not suitable for our artificial GUI design images, especially considering the fact that there is not so much data in this study.
FaceOff is much better than WGAN-GP, but there are still some issues with their approach.
First, FaceOff often chooses the subtrees with the highest back score to accelerate the convergence of the model and only compares the structural similarity between the real GUIs and the retrieved template to minimize their distance, resulting in the diversity loss.
However, it does not consider the relative position of each component especially the specific top-down relationship of the structure in the GUI.
Therefore, most generated GUIs from FaceOff are of very similar structure like that in Fig~\ref{fig:SamplesDisplayBaselines} (b), and many GUIs are also of the same color schema as that in  Fig~\ref{fig:SamplesDisplayBaselines} (c) and Fig~\ref{fig:SamplesDisplayBaselines} (d).

% WGAN-GP and FaceOff

\begin{figure}\scriptsize
    \begin{minipage}{0.25\linewidth}
        \centerline{\includegraphics[width=2.2cm,height=4.4cm]{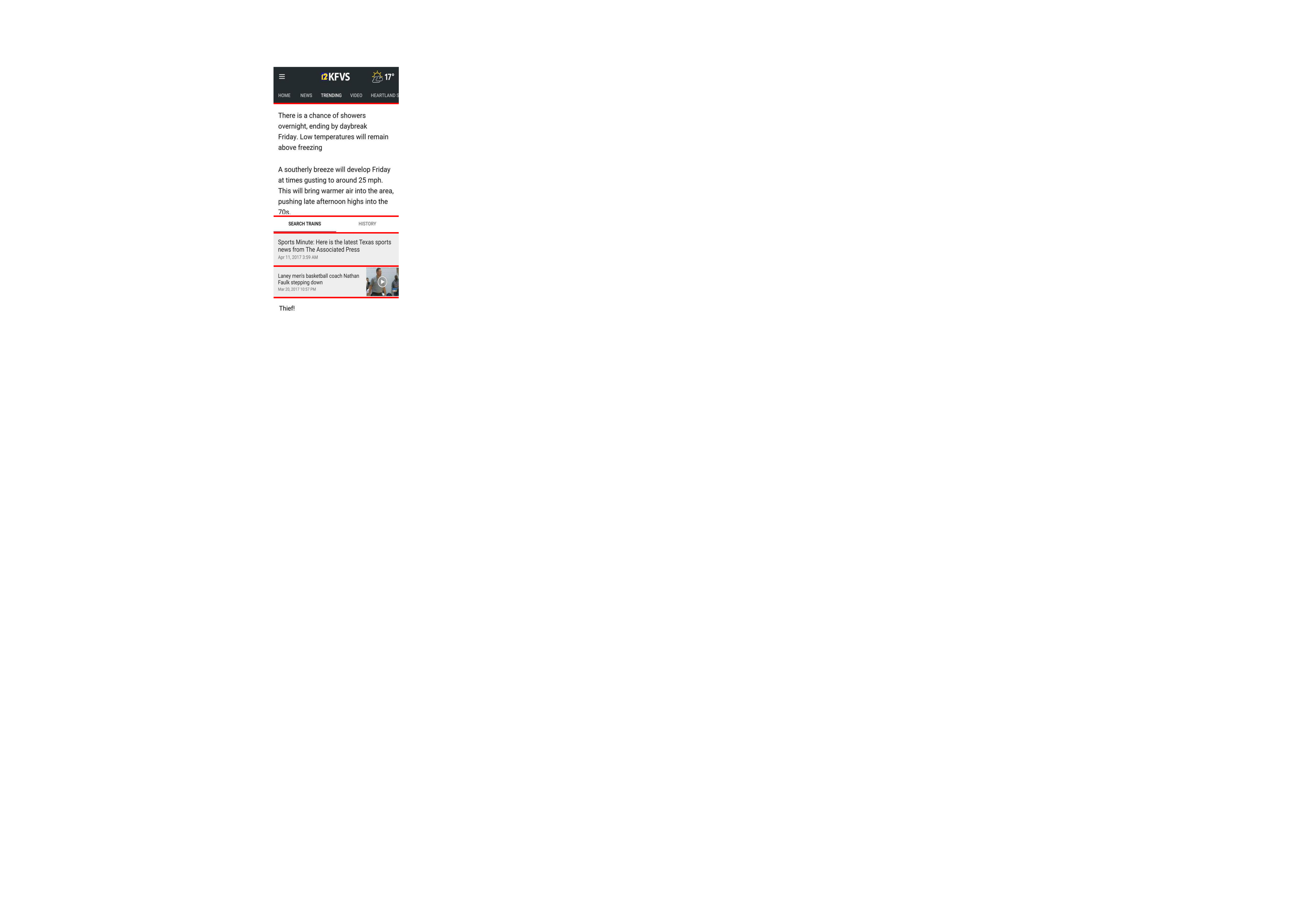}}
        \centerline{(a)}
    \end{minipage}%
    \hfill
    \begin{minipage}{0.25\linewidth}
        \centerline{\includegraphics[width=2.2cm,height=4.4cm]{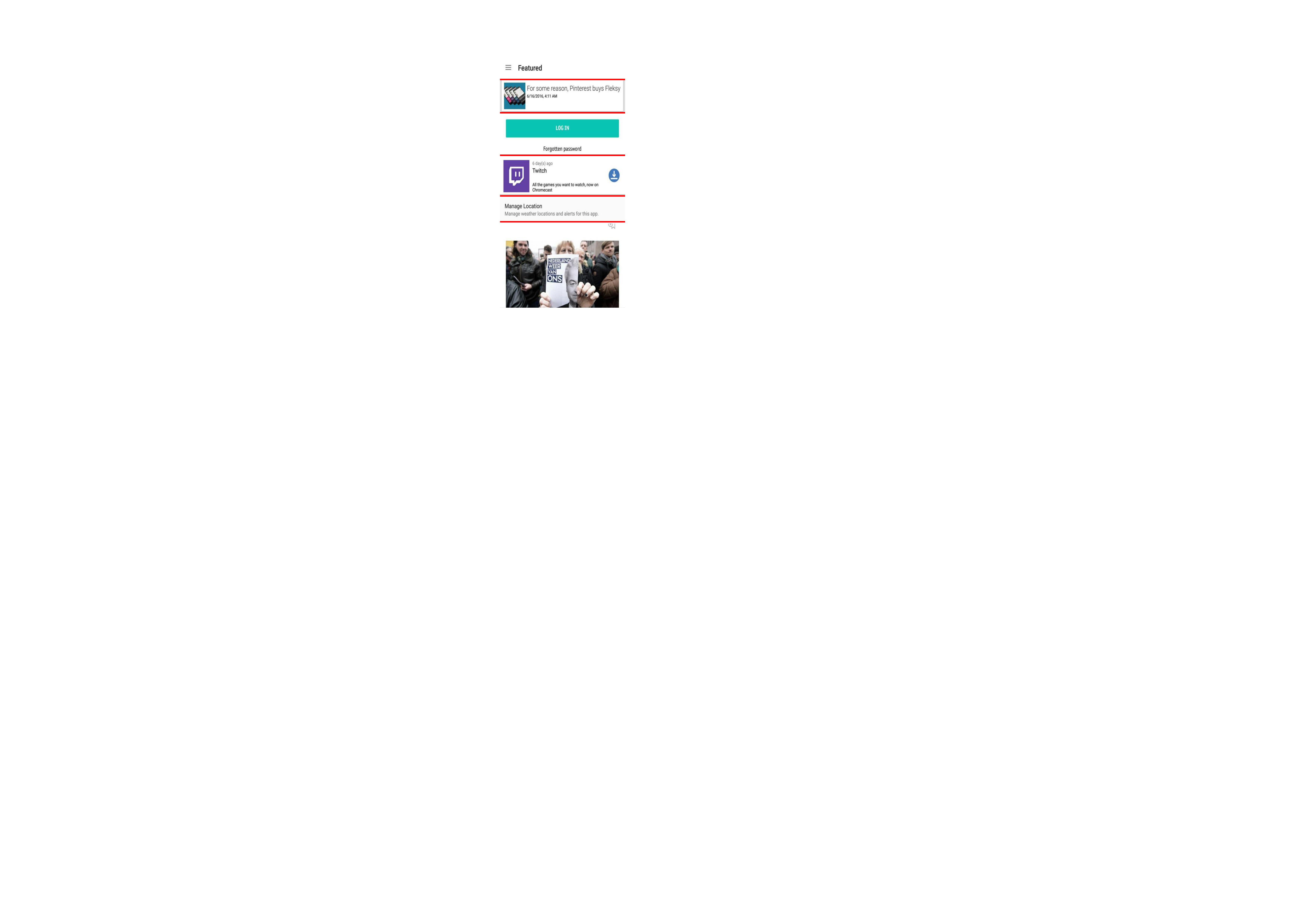}}
        \centerline{(b)}
    \end{minipage}%
    \hfill
    \begin{minipage}{0.25\linewidth}
        \centerline{\includegraphics[width=2.2cm,height=4.4cm]{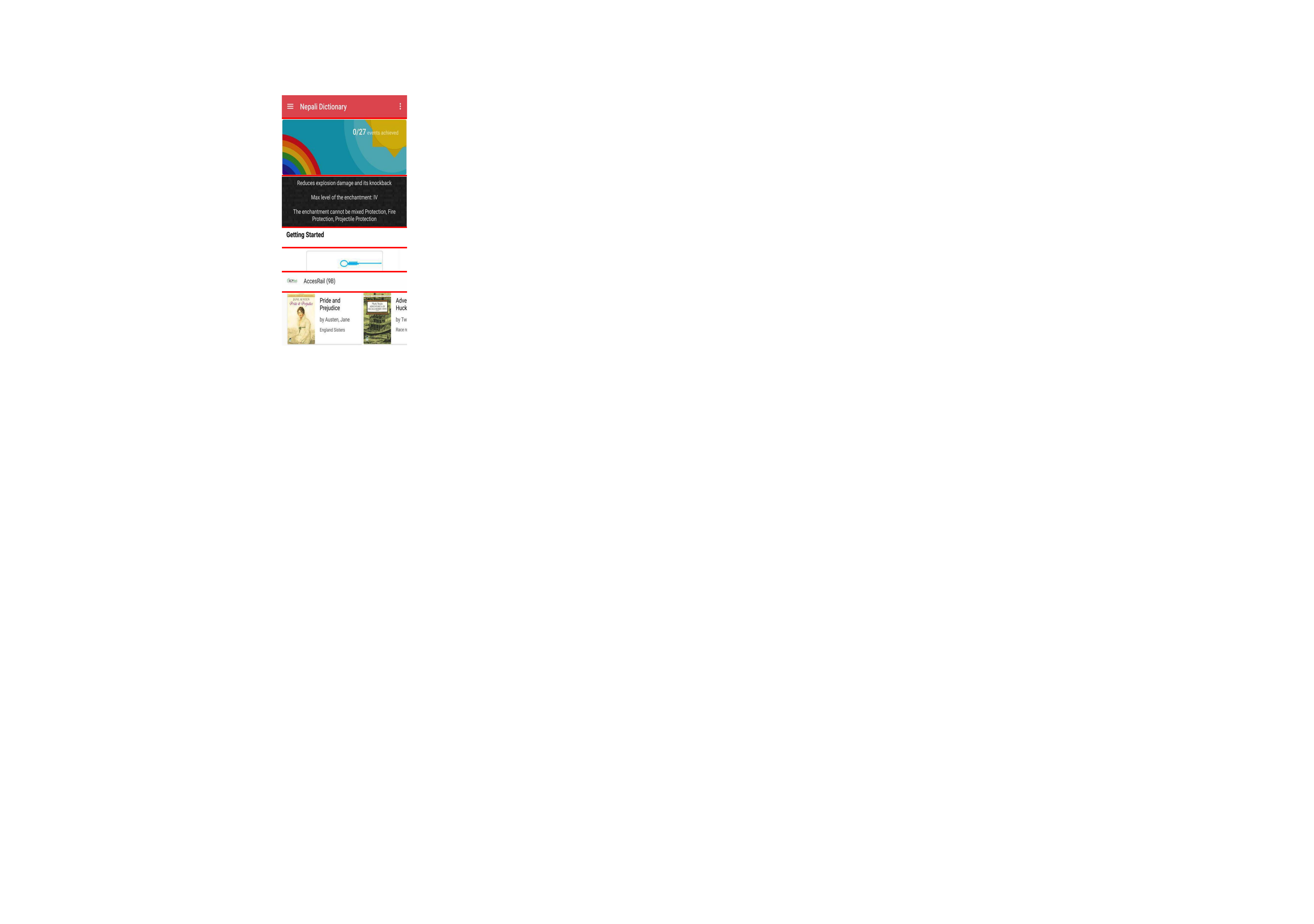}}
        \centerline{(c)}
    \end{minipage}%
    \hfill
    \begin{minipage}{0.25\linewidth}
        \centerline{\includegraphics[width=2.2cm,height=4.4cm]{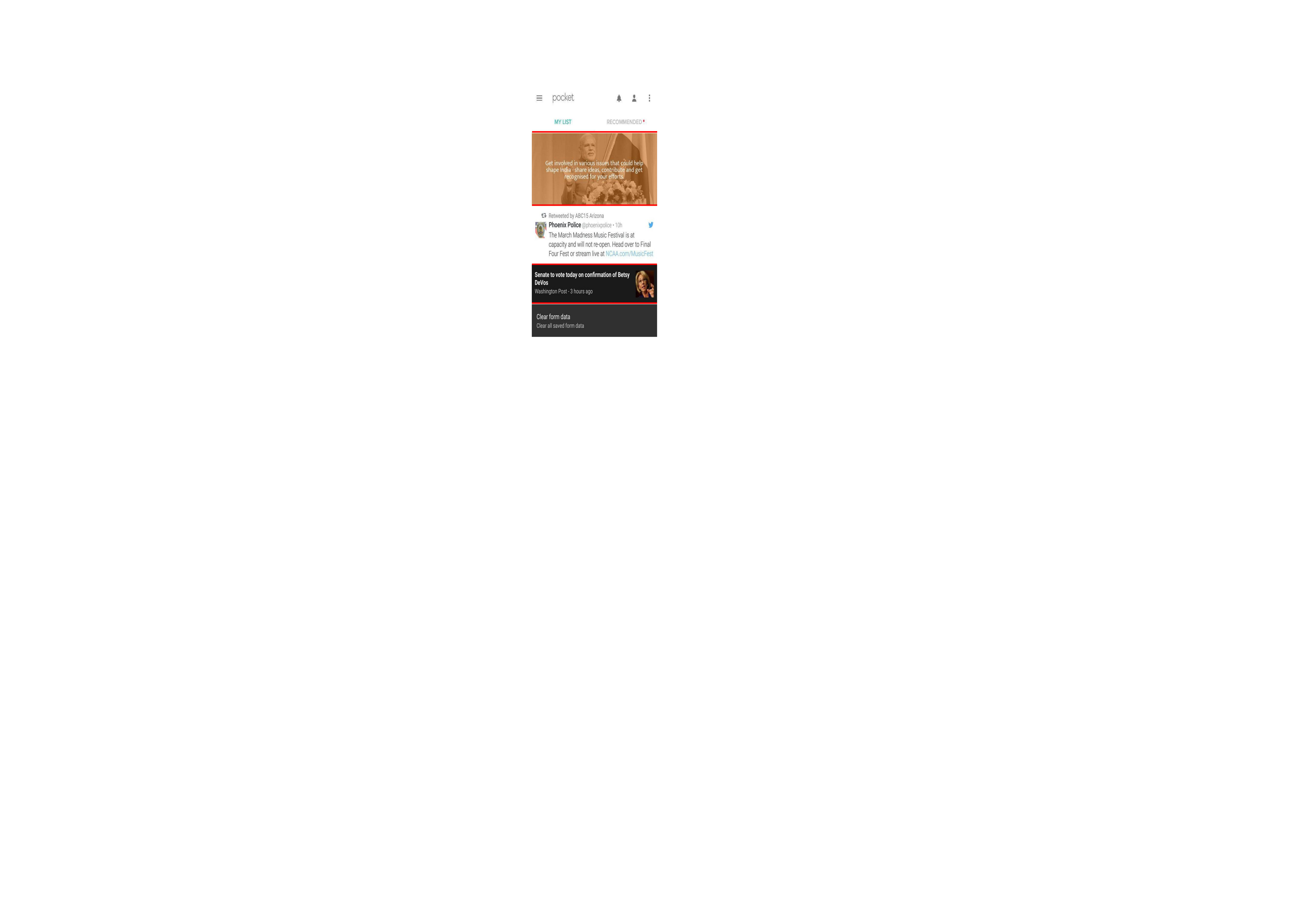}}
        \centerline{(d)}
    \end{minipage}%
    \caption{Generated GUI examples by \tool-style (a, b) and \tool-structure (c, d).}
    \label{fig:SamplesDisplayBaselinesGUIGAN}
\end{figure}

The other two derived baselines of our approach, \tool-style and \tool-structure explore the impact of style information and structure information on the generated results. 
With only modeling the design style information, \tool-style can generate GUI designs with harmonious color combinations as seen in Fig~\ref{fig:SamplesDisplayBaselinesGUIGAN} (a) and (b), but without very good structural designs.
For example, the menu tab appears in the middle of the GUI in Fig~\ref{fig:SamplesDisplayBaselinesGUIGAN} (a) and the login button appears at the top of the GUI in Fig~\ref{fig:SamplesDisplayBaselinesGUIGAN} (b).
Similar issues also apply to \tool-structure with reasonable and diverse layouts of generated GUIs, but terrible color schema as seen in Fig~\ref{fig:SamplesDisplayBaselinesGUIGAN} (c), (d).
The results from these two baselines demonstrate that the two loss function settings in Section~\ref{sec:approach} successfully capture the style and structure information.

% GUIGAN-structure and GUIGAN-style

\begin{figure}[h]\scriptsize
    \begin{minipage}{0.25\linewidth}
        \centerline{\includegraphics[width=2.2cm,height=4.4cm]{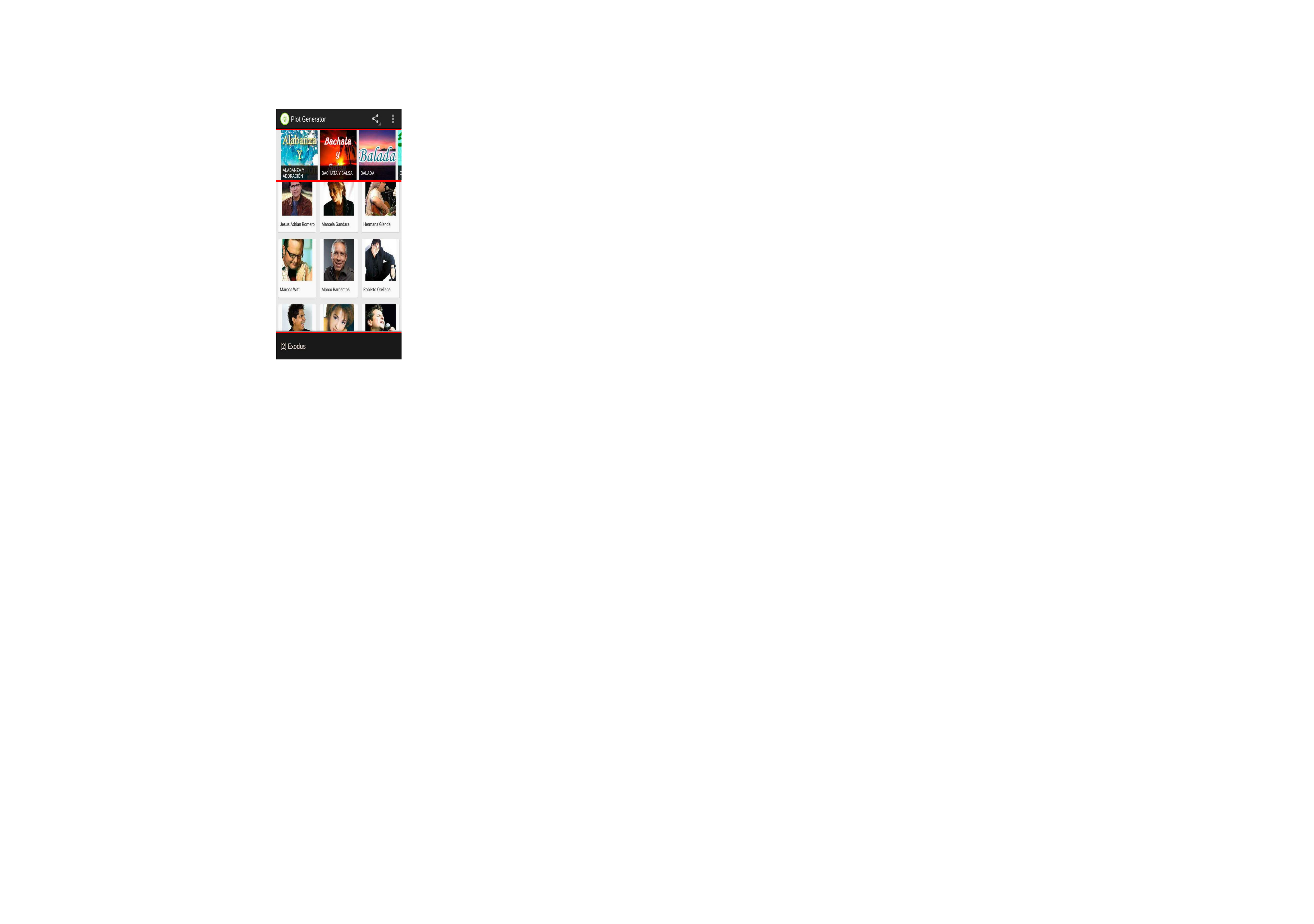}}
        \centerline{(a)}
    \end{minipage}%
    \hfill
    \begin{minipage}{0.25\linewidth}
        \centerline{\includegraphics[width=2.2cm,height=4.4cm]{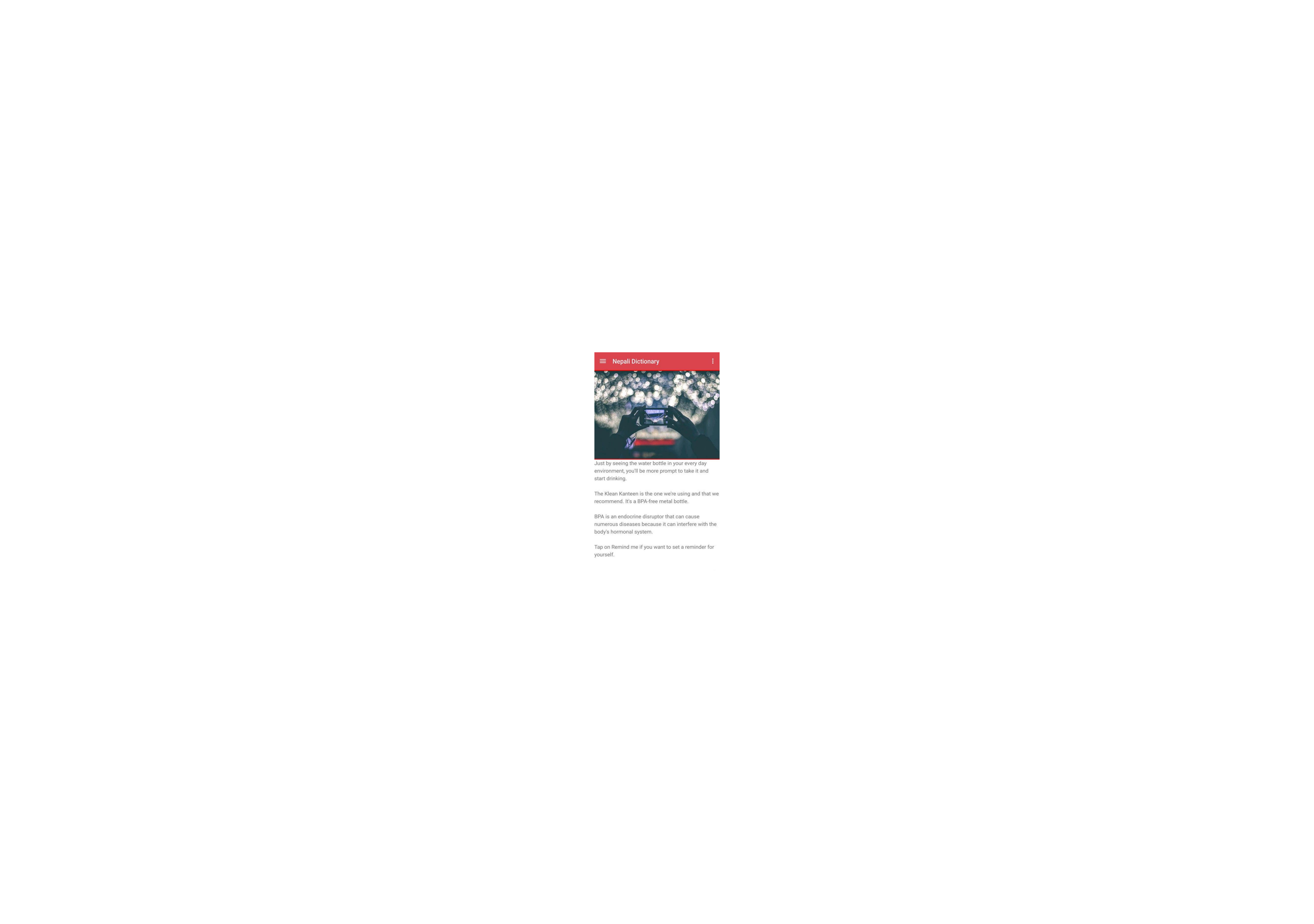}}
        \centerline{(b)}
    \end{minipage}%
    \hfill
    \begin{minipage}{0.25\linewidth}
        \centerline{\includegraphics[width=2.2cm,height=4.4cm]{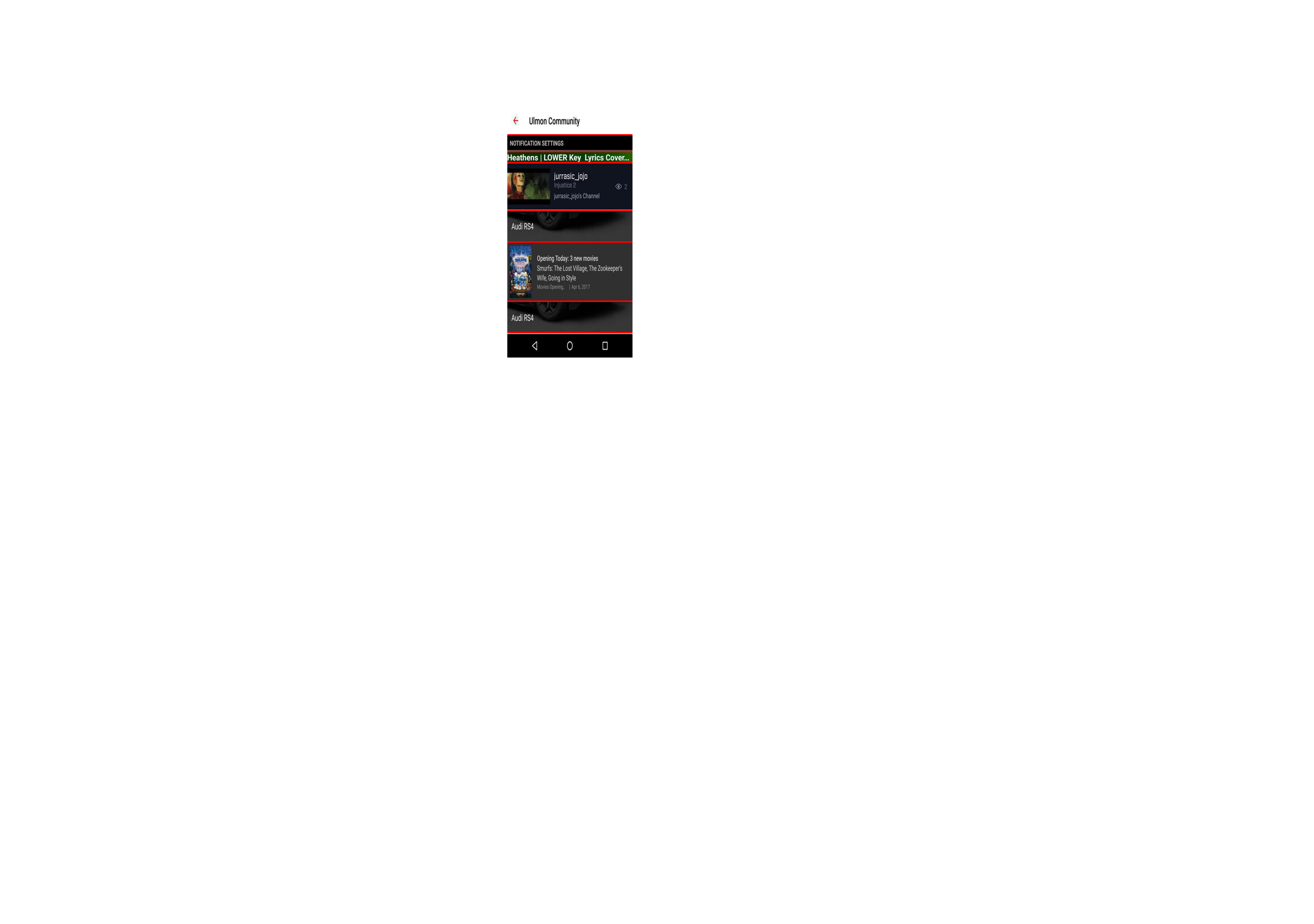}}
        \centerline{(c)}
    \end{minipage}%
    \hfill
    \begin{minipage}{0.25\linewidth}
        \centerline{\includegraphics[width=2.2cm,height=4.4cm]{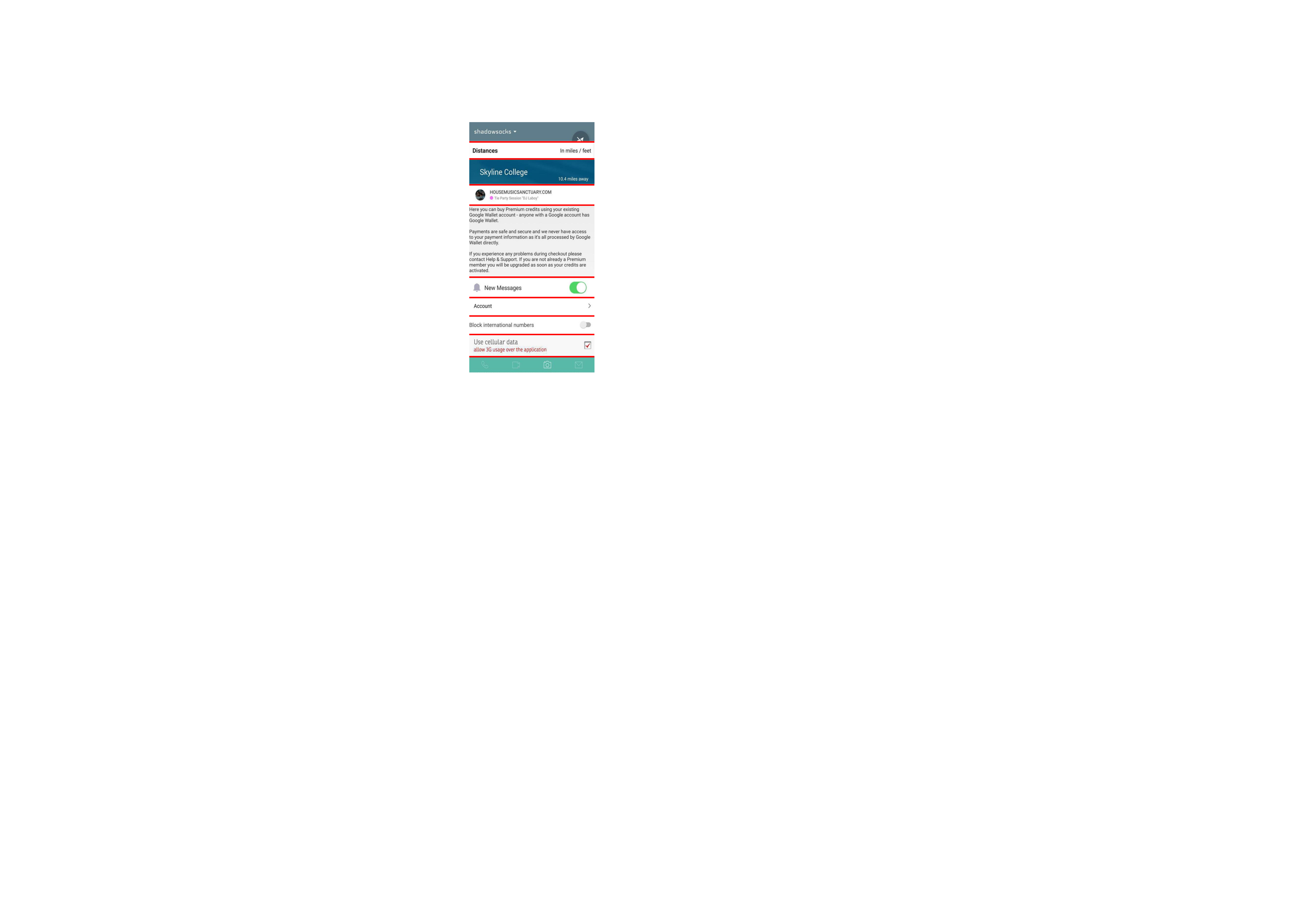}}
        \centerline{(d)}
    \end{minipage}%
    \caption{Examples of bad results generated by \tool.}
    \label{fig:SamplesDisplayGUIGANBad}
\end{figure}

Although most of the samples generated by our model are satisfactory, there are still some bad designs. 
We manually observe those bad GUI designs and summarize some reasons.
First, due to the default size of components in the subtree, some of them are difficult to fit into the generated GUIs as seen in Fig~\ref{fig:SamplesDisplayGUIGANBad} (a), (b).
There is either an overlap between some components or one figure taking all the GUI space.
Second, since our model learns the style and structure information at the same time, there may be an imbalance between them for some GUI generation.
Fig~\ref{fig:SamplesDisplayGUIGANBad} (c) shows an example with too much emphasis on style consistency while ignoring the structural effects.
In contrast, Fig~\ref{fig:SamplesDisplayGUIGANBad} (d) has a set of diverse components in good structure but incompatible color schema.

% GUIGAN bad

\section{Human Evaluation}
The target of this work is to automatically generate a list of GUI designs for novice designers or developers to adopt. 
The automated experiments above demonstrate the performance of our model compared with other baselines.
However, the satisfactoriness of the GUI design can be subjective depending on different users or developers.
To better evaluate the usefulness of \tool, we conduct a user study to investigate the feedback from developers in this section.

\subsection{Evaluation Metrics}
There are no existing evaluation metrics for mobile GUI design in the literature. 
But inspired by the web GUI evaluation~\cite{reinecke2013predicting, coursaris2008empirical, 2020Attribute} and image evaluation~\cite{zhang2008image, wang2020use}, we propose three novel metrics for participants to rate the quality of the GUI design from three aspects by considering the characteristics of the mobile GUIs.
First, design aesthetics is to evaluate the overall design's pleasing qualities.
Second, we adopt color harmony~\cite{tokumaru2002color, yang2016automatic} which refers to the property that certain aesthetically pleasing color combinations have to evaluate the color schema selection within the GUI.
These combinations create pleasing contrasts and consonances that are said to be harmonious.
Third, structure rationality is used to \revised{measure the component layout rationality, i.e., the location of components in the GUI and the logic of their combination and sorting.}
For each metric, the participants will give a score ranging from 1 to 5 with 1 representing the least satisfactoriness while 5 as the highest satisfactoriness.
\revised{Besides, to confirm if \tool implicitly considers the app functionality during the training, we further ask participants to manually check if the component distribution of generated GUIs is functionally correct e.g., the menu bar on the top of the page.
They will mark 1 if the GUI components are functionally correctly distributed, while 0 for incorrect ones.}

\subsection{Procedures}
In real-world app development, teams often know their target very well.
To mimic that practice, we select 5 app categories (same in Section~\ref{sec:auto-evaluation}) for specific GUI generation.
For each category, we randomly generate 10 GUI designs for each method.
Due to the poor performance of WGAN-GP in the last experiment, we only take FaceOff as the baseline.

We recruited five Master students majoring in computer science.
They all have several-year programming experience and at least 1-year Android development experience mostly about GUI implementation and some GUI design.
Therefore, they can be regarded as junior Android developers for evaluating whether they are satisfied with our GUI design.
First, we introduce them a detailed explanation about the GUI evaluation metrics.
Then they are provided with the generated GUI designs from different methods, then give the score of each GUI design in three metrics i.e., design aesthetics, color harmony, structure rationality.
Note that they do not know which GUI design is from which method and all of them will evaluate the GUI design individually without any discussion.
After the experiment, we tell the participants which GUI designs are generated by our model and ask them to leave some general comments about our \tool.

\subsection{Results}

\renewcommand{\arraystretch}{1.5} %Ã¦ÂÂ§Ã¥ÂÂ¶Ã¨Â¡ÂÃ©Â«Â  
\renewcommand{\multirowsetup}{\centering}
\begin{table}
  \centering  
  \fontsize{8}{7}\selectfont  
  \begin{threeparttable}  
  \caption{Performance of Human Evaluation. ** denotes $p < 0.01$ and * denotes $p < 0.05$}
    \begin{tabular}{ccll} 
    \toprule  
    \multirow{2}{1in}{Category}&  \multirow{2}{0.6in}{Metric}&\multicolumn{2}{c}{Score}\cr
    %{Real-World}&{FaceOff}&\multicolumn{3}{c}{\tool}\cr
    \cmidrule(lr){3-4}
    &&FaceOff&\tool\cr  
    \midrule  
    \multirow{4}{1in}{News \& Magazines}
    & aesthetics &$2.08$&$2.96^{**}$\\
    & harmony &$2.54$&$3.18^{**}$ \\
    & structure &$2.22$ &$3.02^{**}$ \\
    & functionality&0.38&$0.82^{**}$\cr
    \cmidrule(lr){1-4}
    \multirow{4}{1in}{Books \& Reference} 
    & aesthetics & $2.40$ & $3.16^{**}$ \\
    & harmony & $2.46$ & $3.32^{**}$ \\
    & structure &  $2.40$ & $3.40^{**}$ \\
    & functionality&0.40&$0.74^{**}$\cr
    \cmidrule(lr){1-4}
    \multirow{4}{1in}{Shopping} 
    & aesthetics &2.66 &3.02 \\
    & harmony &3.04 &3.18 \\
    & structure &$2.52$ &$3.00^{*}$ \\
    & functionality&0.60&0.78\cr
    \cmidrule(lr){1-4}
    \multirow{4}{1in}{Communication} 
    & aesthetics &$2.56$ &$3.12^{*}$ \\
    & harmony &$2.86$ &$3.42^{*}$ \\
    & structure &$2.60$ &$3.18^{*}$ \\
    & functionality&0.42&$0.82^{**}$\cr
    \cmidrule(lr){1-4}
    \multirow{4}{1in}{Travel \& Local}
    & aesthetics  &$2.14$ &$3.30^{**}$ \\
    & harmony &$2.30$ &$3.38^{**}$ \\
    & structure  &$2.16$ &$3.44^{**}$ \\
    & functionality&0.46&$0.90^{**}$\cr
    \cmidrule(lr){1-4}
    \multirow{4}{1in}{\textbf{Average}}
    & aesthetics &$2.37$ & $3.11^{**}$\\
    & harmony  &$2.64$ &$3.30^{**}$ \\
    & structure &$2.38$ &$3.21^{**}$ \\
    & functionality&$0.452$ &$0.812^{**}$\cr
    \bottomrule  
    \end{tabular}  
    \label{tab:HumanEvaluation}
    \end{threeparttable} 
\end{table}

As shown in Table~\ref{tab:HumanEvaluation}, the generated GUI designs from our model outperforms that of FaceOff significantly with 3.11, 3.30, and 3.21 which are 31.22\%, 25.00\%, and 34.87\% increase in overall aesthetics, color harmony, and structure.
In addition to the average score, our model is also better than FaceOff in generating GUI design for all five app categories in three metrics.
That result also demonstrates the generalization of our \tool.
According to the detailed analysis of the experiment result, the GUI designs with low scores tend to have incomplete structure, single content, large and abrupt pictures, or advertisements. 
In contrast, the GUI with high scores has a concise layout, slightly rich content, and background compatible images. We also find that users' requirements for content richness are much higher than other indicators, but this often goes against the simplicity of layout, which needs further research and balance.
 
To understand the significance of the differences between the two approaches, we carry out the Mann-Whitney U test~\cite{fay2010wilcoxon} which is specifically designed for small samples (only 10 GUI design in each category) on three metrics.
The test results in Table~\ref{tab:HumanEvaluation} suggests that our \tool can contribute significantly to the GUI design in all three metrics  with $p-value < 0.01$ or $p-value < 0.05$ except the aesthetics and color harmony metrics in the shopping category.

\begin{figure}
    \begin{minipage}{0.5\linewidth}
        \centerline{\includegraphics[width=0.9\linewidth]{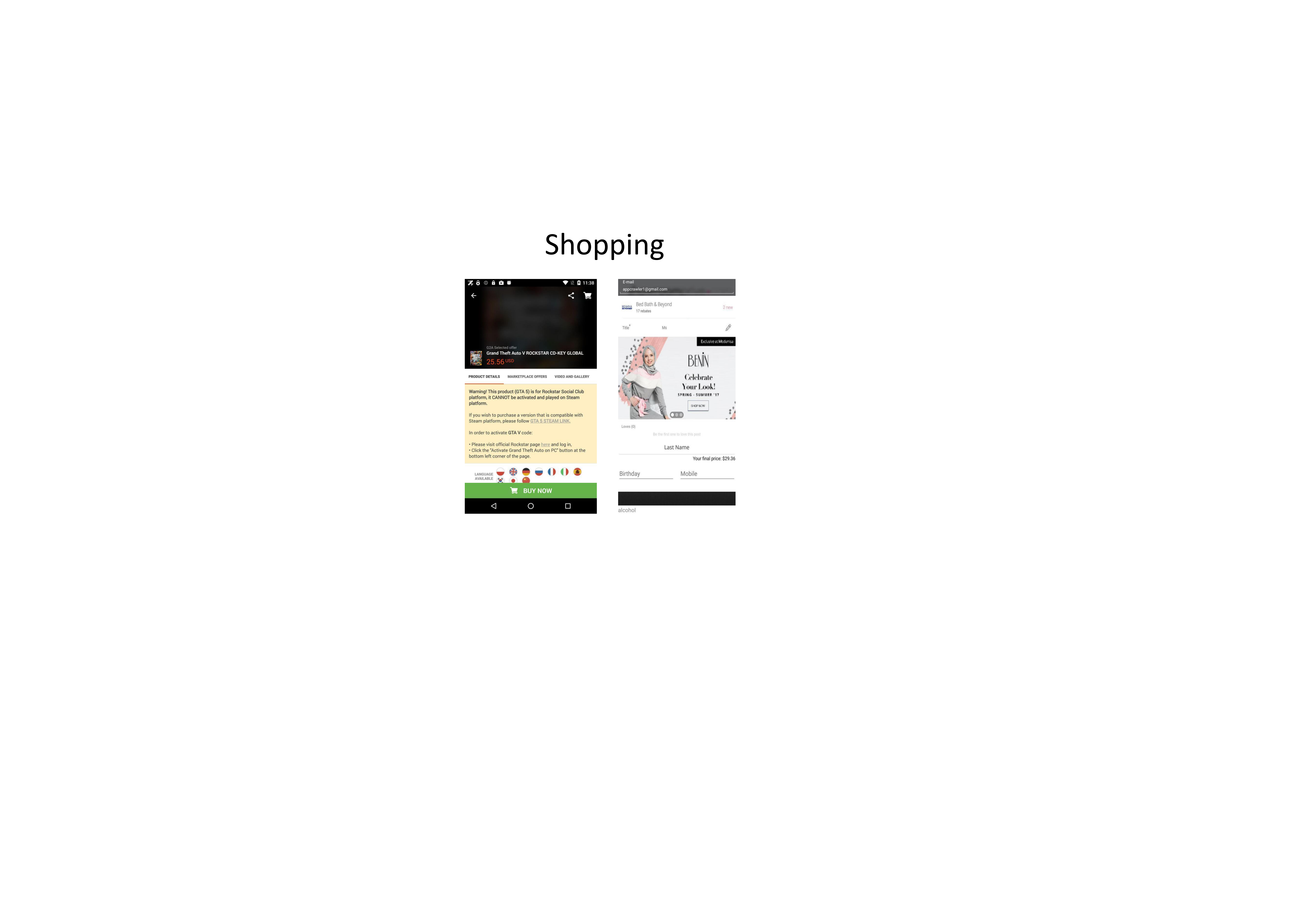}}
        \centerline{(a)~Shopping}
    \end{minipage}%
    \hfill
    \begin{minipage}{0.5\linewidth}
        \centerline{\includegraphics[width=0.9\linewidth]{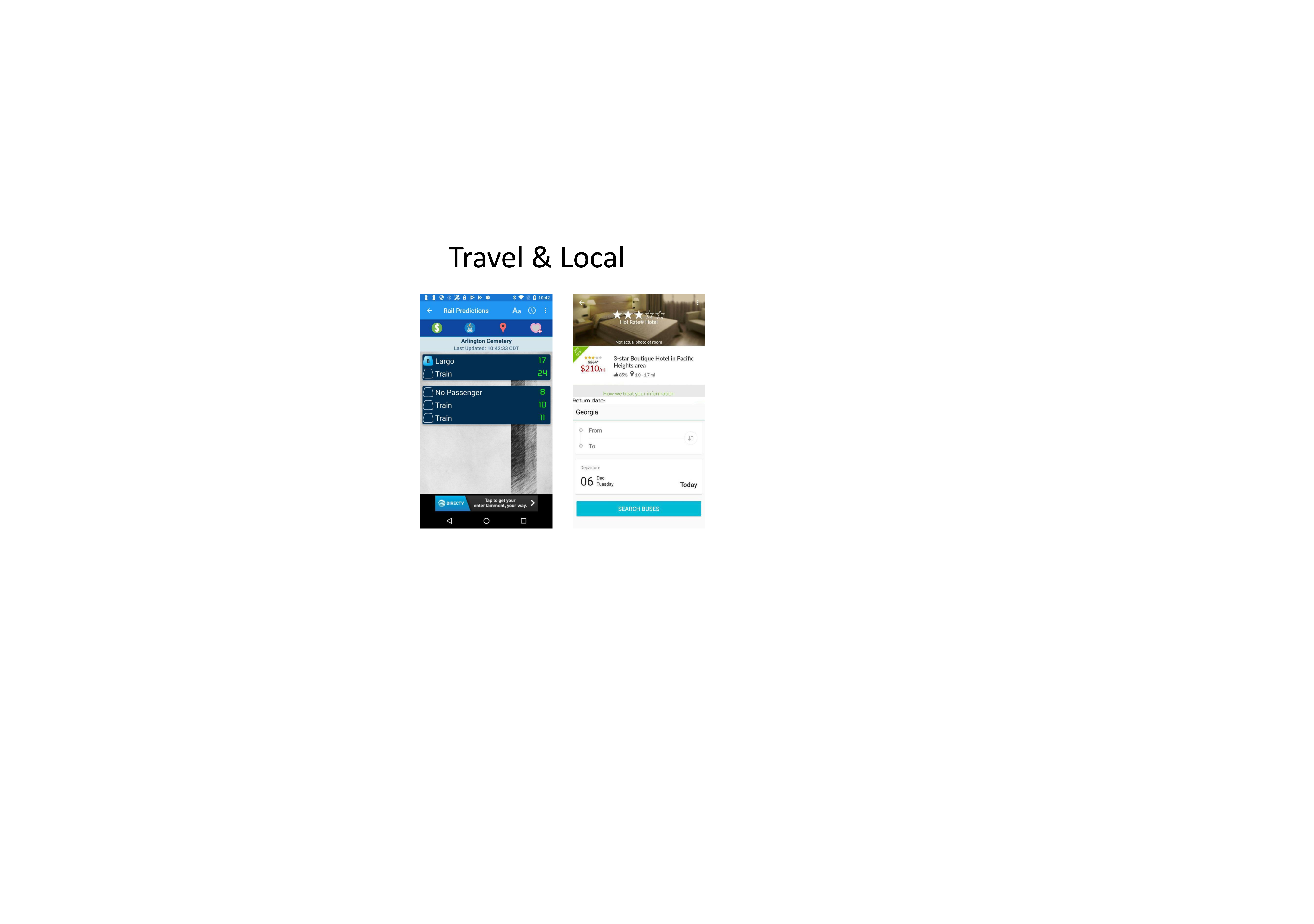}}
        \centerline{(b)~Travel \& Local}
    \end{minipage}%
    \caption{In each pair, the first image is real-world GUI while the second one is generated by \tool.}
    \label{fig:BetterCases}
\end{figure}

Besides the comparison with the baseline, we also present participants another dataset mixing with 10 randomly selected real-world GUI design images from our dataset and 10 randomly selected GUI designs generated by our model for checking the overall GUI aesthetics, color harmony, and structure.
\revised{Some  generated GUIs are even rated higher than real-world ones. In terms of Fig~\ref{fig:BetterCases} (a), the five user-study participants rate the real GUI with 2.8, 3.2, and 3.6 in three metrics (aesthetics, harmony, structure) on average, while 3, 3.4, 3.4 for the generated one. For Fig~\ref{fig:BetterCases} (b), they score the real GUI with 3.2, 3.6, and 4, and the generated GUI with 3.6, 3.6, and 4 which is not that high. There are several reasons why our generated GUIs get higher scores than some real-world GUIs: (1) Some generated GUIs (examples in Fig~\ref{fig:BetterCases}) are of higher quality than some poorly designed real-world GUIs. And note that the score is just 2.1\% to 3.7\% higher than some real-world GUIs. (2) There may be the human bias of different raters as different people are of different aesthetic values. Different human raters may also adopt slightly different criteria when rating the GUI quality as people vary. We mitigated those potential bias by not telling them which GUI is generated by our model or from real-world apps.}

\revised{The results of functionality can be seen in Table IV with the average score as 0.812 which is 79.65\% significantly higher than that of the baseline. It shows that the component distribution of most of our generated GUIs is functionally correct. It also indicates our model implicitly get that app functionality during training on a large-scale dataset as the GUI design is highly related to app functionality, though our model does not explicitly consider app functionality. Additionally, we find that the performance of our model differs in different app categories. \tool achieves an average score of 0.9 in the Travel \& Local category, but only 0.74 in the Books \& Reference category, which may be because the former one is of more functional characteristics than the latter one.}

After the experiment, we tell participants which GUIs are generated from our tool and get some feedback from them. 
They suggest that the padding between components should be adjusted to increase the attractiveness of the GUI designs. 
In addition, they also request to add new features that allow manual settings of result diversity, such as a comparison between combinations, style, structure, etc.

\section{Related Work}
GUI is crucial for the user experience of modern desktop software, mobile applications, and online websites. 
In this section, we introduce related works about GUI design and GUI generation.

\subsection{GUI Design}
GUI design is an important step in GUI development.
Therefore, many researchers are working on assisting designers in the GUI design such as investigating the UI design patterns~\cite{alharbi2015collect}, color evolution~\cite{jahanian2017colors, jahanian2017mining}, UI-related users' review~\cite{fu2013people, martin2017survey}, GUI code generation~\cite{chen2018ui}, website GUI generation~\cite{zheng2019faceoff}, etc.
Liu et al.~\cite{liu2018learning} follow the design rules from Material Design to annotate the mobile GUI design to represent its semantics.
Swearngin et al.~\cite{swearngin2018rewire} adopt the image processing method to help designs with converting the mobile UI screenshots into editable files in Photoshop so that designers can take it as a starting point for further customization. 
To render inspirations to the designer, Chen et al.~\cite{chen2019storyboard} propose a program-analysis method to efficiently generate the storyboard with UI screenshots, given one app executable file. 
Fischer et al.~\cite{fischerimaginenet} transfer the style from fine art to GUI. 
Chen et al.~\cite{chen2020object} study different GUI element detection methods on large-scale GUI data and develop UIED~\cite{2020UIED} to handle diverse and complicated GUI images. Other supporting works such as GUI tag prediction~\cite{chen2020lost} and GUI component gallery construction~\cite{chen2019gallery} can enhance designers' searching efficiency.

All of these works are targeting at simplifying the design process for professional designers. In contrast, our method is focusing on the initial stage of GUI design i.e., generating diverse GUI designs for giving inspirations to novice designers and developers who are of not much GUI design training. 
Through the GAN method in deep learning, our model learns the design style and structural characteristics of existing GUIs to generate diversified new GUI for designers' reference, so as to lower the GUI design barrier.

\subsection{GUI Generation}
Thanks to the rapid development of deep learning, the image generation performance are further improved, especially by the Generative Adversarial Network (GAN)~\cite{goodfellow2014generative} and its deviation models. 
Apart from the natural image generation, there are also many works on re-arranging elements for composing graphic designs with better layouts (especially semantic layouts).

Sandhaus et al.~\cite{sandhaus2011employing} present an approach for the automatic layout of photo compositions that incorporates the knowledge about aesthetic design principles. Yang et al.~\cite{yang2016automatic} analyze low-level image features and apply high-level aesthetic designing principles and predefined templates to the given images and texts, thus to automatically suggest the optimal template, text locations, and colors. 
Vempati et al.~\cite{vempati2019enabling} utilize the MaskRCNN object detector to automatically annotate the required objects/tags and a Genetic Algorithm method to generate an optimal advertisement layout for the given image content, input components, and other design constraints. A ranking model is trained on historical banners to rank the generated captivities by predicting their Click-Through-Rate (CTR).
Li et al.~\cite{li2019controllable} introduce a progressive generative model of image extrapolation with three stages and two important
sub-tasks.

In the field of text-to-image synthesis, Hinz et al.~\cite{hinz2019semantic} introduce Semantic Object Accuracy (SOA) to evaluate images given an image caption.
LayoutGAN is proposed by Li et al.~\cite{li2020layoutgan} for graphic design and scene generation, introducing wireframe rendering for image discrimination. The generator takes as input a set of vectors and uses self-attention modules to refine their labels and geometric parameters jointly. Jyothi et al.~\cite{jyothi2019layoutvae} propose a variational autoencoder based method called LayoutVAE, which allows for generating full image layouts given a label set, or per label layouts for an existing image given a new label and has the capability of detecting unusual layouts.
Some works are generating GUI test case for checking GUI usability~\cite{zhao2020seenomaly, bo2021don, liu2020owl}, accessibility~\cite{chen2020unblind} and security~\cite{chen2019gui}.

Unlike these works in generating the layout of graphic design like posters, advertisements, we are the first to work on GUI design generation.
Different from their tasks in arranging the given components, our task is more challenging i.e., selecting components from a repository and compose them into a great GUI design by taking the design style and structure information into the consideration.
Therefore, we develop a novel approach for modeling that information.

\section{Conclusion}
Designing a good GUI which requires much innovation and creativity is difficult even to well-trained designers.
In this paper, we propose a GAN-based GUI design generation method, \tool, which can assist novice designers and developers by generating new GUIs by learning the existing app GUI screenshots.
The generated GUI design can be regarded as the starting point or inspiration for their design work. 
We decomposed the filtered GUIs to form our large-scale subtree retrieval repository, then feed these subtrees to our model for generating reasonable one-dimension sequences, which are further used for reorganization. 
Two additional corrections are added to the generator of our model to improve the model in the aspect of design style and structural composition. 
The automated experiments demonstrate the performance of our model and the user study confirms the usefulness of {\tool}.

\revised{To improve the generated GUIs, we will adopt two ways. First, we will improve our model to generate GUIs with higher quality. We can summarize a list of issues of the current model by carrying out a detailed analysis of a bad GUI generation of current data. We then improve our model accordingly and also add a list of rules to post process generated GUIs e.g., menu bar should be on the top of the GUI, etc. Second, we will build an AI-human collaboration system i.e., the generated GUIs in our model are only used for inspiring developers/designers. Designers or developers can further select or customize GUIs according to their purpose. Besides, our model can be used to convert designers’/developers’ partial GUI design to the full one via our model’s leveraging the pre-built GUI components and control, though it can be further improved. Some examples can be seen at our online gallery\footnote{\url{https://github.com/GUIDesignResearch/GUIGAN/blob/master/README.md\#examples-of-pre-built-components}}.}
We also plan to combine the current GUI code generation works~\cite{chen2018ui, moran2018machine, sidong2021auto} with our GUI design generation to fully automate GUI development.

\section{Acknowledgements}
This work is supported in part by the National Natural Science Foundation of China (Grant 61471181), and Chunyang Chen is partially supported by Facebook research award.

\bibliographystyle{IEEEtran}
\normalem
\bibliography{ICSE2021}

\end{document}